\documentclass[iop,apj,twocolappendix]{emulateapj}
\usepackage{amsmath,amssymb,amstext}

\usepackage[breaklinks,colorlinks,citecolor=blue,linkcolor=magenta]{hyperref} 
\usepackage[all]{hypcap} 
\usepackage{multirow}
\usepackage{booktabs}
\usepackage{enumitem}
\usepackage{rotating}

\usepackage{aas_macros}
\usepackage{natbib}
\shorttitle{Star Formation History of NGC~7793}

\shortauthors{Sacchi et al.}

\begin{document}

\title{Star Formation Histories of the LEGUS Spiral Galaxies.\\I. The flocculent spiral NGC~7793 \footnotemark[$\star$]} \footnotetext[$\star$]{Based on observations obtained with the NASA/ESA \textit{Hubble Space Telescope} at the Space Telescope Science Institute, which is operated by the Association of Universities for Research in Astronomy under NASA Contract NAS 5-26555.}
\author{E. Sacchi$^{1}$, M. Cignoni$^{2,3,4}$, A. Aloisi$^{1}$, M. Tosi$^{4}$,
A. Adamo$^{5}$,
D. A. Dale$^{6}$,
B. G. Elmegreen$^{7}$,
D. M. Elmegreen$^{8}$,
D. Calzetti$^{9}$,
D. A. Gouliermis$^{10,11}$,
K. Grasha$^{12}$,
L. J. Smith$^{13}$,
A. Wofford$^{14}$,
J. C. Lee$^{15,16}$,
E. Sabbi$^{1}$, and
L. Ubeda$^{1}$
}
\affil{
$^{1}$Space Telescope Science Institute, 3700 San Martin Drive, Baltimore, MD 21218, USA; esacchi@stsci.edu\\
$^{2}$Dipartimento di Fisica, Universit\`a di Pisa, Largo Bruno Pontecorvo, 3, 56127 Pisa, Italy\\
$^{3}$INFN, Sezione di Pisa, Largo Pontecorvo 3, 56127 Pisa, Italy\\
$^{4}$INAF--Osservatorio di Astrofisica e Scienza dello Spazio di Bologna, Via Gobetti 93/3, I-40129 Bologna, Italy\\
$^{5}$Dept. of Astronomy, The Oskar Klein Centre, Stockholm University, Stockholm, Sweden \\
$^{6}$Department of Physics and Astronomy, University of Wyoming, Laramie, WY\\
$^{7}$IBM Research Division, T.J. Watson Research Center, Yorktown Hts., NY\\
$^{8}$Department of Physics and Astronomy, Vassar College, Poughkeepsie, NY\\
$^{9}$Department of Astronomy, University of Massachusetts -- Amherst, Amherst, MA 01003, USA\\
$^{10}$Zentrum f\"ur Astronomie der Universit\"at Heidelberg, Institut f\"ur Theoretische Astrophysik, 69120 Heidelberg, Germany\\
$^{11}$Max Planck Institute for Astronomy,  K\"{o}nigstuhl\,17, 69117 Heidelberg, Germany\\
$^{12}$Research School of Astronomy and Astrophysics, The Australian National University, Weston Creek, ACT 2611, Australia\\
$^{13}$European Space Agency/Space Telescope Science Institute, Baltimore, MD\\
$^{14}$Instituto de Astronom\'ia, Universidad Nacional Aut\'onoma de M\'exico, Unidad Acad\'emica en Ensenada, Km 103 Carr. Tijuana-Ensenada, Ensenada 22860, M\'exico\\
$^{15}$Department of Physics \& Astronomy, California Institute of Technology, Pasadena, CA, USA\\
$^{16}$Infrared Processing and Analysis Center, California Institute of Technology, Pasadena, CA, USA
}

\begin{abstract}
We present a detailed study of the flocculent spiral galaxy NGC~7793, part of the Sculptor group. By analyzing the resolved stellar populations of the galaxy, located at a distance of $\sim 3.7$~Mpc, we infer for the first time its radial star formation history (SFH) from \textit{Hubble Space Telescope} photometry, thanks to both archival and new data from the Legacy ExtraGalactic UV Survey. We determine an average star formation rate (SFR) for the galaxy portion covered by our F555W and F814W data of $0.23\pm0.02$~$\mathrm{M_{\odot}/yr}$ over the whole Hubble time, corresponding to a total stellar mass of $(3.09\pm0.33)\times 10^9$~$\mathrm{M_{\odot}}$ in agreement with previous determinations. Thanks to the new data extending to the F336W band, we are able to analyze the youngest stellar populations with a higher time resolution. Most importantly, we recover the resolved SFH in different radial regions of the galaxy; this shows an indication of a growing trend of the present-to-past SFR ratio, increasing from internal to more external regions, supporting previous findings of the inside-out growth of the galaxy.
\end{abstract}


\maketitle

\section{Introduction}
\setcounter{footnote}{16}
\urlstyle{sf}
Star formation (SF) is a very complex process, that can be regulated by many aspects of stellar, gas and galactic evolution. The connection between the various regimes in which SF occurs is still poorly explored, as well as the transition between different physical scales that interest the SF process, such as the accretion of gas onto disks from satellite objects and the intergalactic medium, the cooling of this gas to form a cool neutral phase, the formation of molecular clouds, the fragmentation and accretion of this molecular gas to form progressively denser structures such as clumps and cores, and the subsequent contraction of the cores to form stars and planets \citep{Kennicutt2012}.

Moreover, in galaxies of different morphological types, SF can proceed in completely different ways, from the giant elliptical galaxies whose SF peaked billions of years ago and is now quenched, to the star forming and starburst galaxies that evolved more slowly and have still a very prolific activity \citep{Brinchmann2004}.

Even among star forming galaxies only, we want to explore the differences and/or similarities among spiral, irregular and dwarf galaxies, and whether we can trace a smooth transition from one another.

In this context, constraining and understanding the star formation histories (SFHs) of individual galaxies is fundamental to unveil the mechanisms that regulate their evolution and lead to the objects we see today.

In this paper, we present the analysis of NGC~7793, a flocculent spiral galaxy (morphological type SA(s)d) part of the Sculptor group, located at a distance of $3.7 \pm 0.1$~Mpc \citep{Radburn-Smith2011,Sabbi2018}. NGC~7793 is a typical late-type Sd galaxy, with a very tiny bulge and a filamentary spiral structure; its stellar mass is $M_{\ast} \sim 3.32 \times 10^9$~M$_{\odot}$ \citep{Dale2009} while the neutral gas mass is $M_{\mathrm{H}\, \textsc{i}} \sim 6.8 \times 10^8$~M$_{\odot}$ \citep{Walter2008}. The Sculptor group has the interesting characteristics of being at a high galactic latitude ($-77 \fdg 2$), that minimizes foreground extinction and contamination from the Milky Way, of containing mainly isolated disk systems, and of being at a distance that allows resolved stellar population studies of its member galaxies (it is indeed the closest group of galaxies beyond the Local Group). The disk of NGC~7793 is particularly well studied, since it exhibits a discontinuity in the metallicity gradient which becomes positive in the outer disk \citep{Vlajic2011}. Also, the radial surface brightness profiles of different stellar populations have a disk break that appears at a constant radius for all stellar ages, but older stars show a steeper profile internal to the break and a shallower profile beyond the break in comparison to younger stars, indicative of high levels of stellar radial migration \citep{Radburn-Smith2012} and/or past accretion events \citep{Abadi2006}.

\citet{Bibby2010} studied the population of Wolf-Rayet (WR) stars in NGC~7793, i.e. helium burning stars descendants of massive O stars with very strong stellar winds. They find 52 of these sources, and with additional slits on the H~\textsc{ii} regions, they estimate a metallicity gradient using strong line calibrations of $12 + \log(O/H) = 8.61 \pm 0.05 - (0.36 \pm 0.10)\ r/R_{25}$ and a SFR of $0.45$~M$_{\odot}$ yr$^{-1}$. Another study by \citet{Stanghellini2015} analyzed the strong-line oxygen abundances in the H~\textsc{ii} regions of the galaxy, finding similar radial metallicity gradients, inside $R_{25} = 5.24 \pm 0.24$ arcmin, also in agreement with the pioneering study by \citet{Edmunds1984}.

\begin{figure*}
\centering
\includegraphics[width=\linewidth]{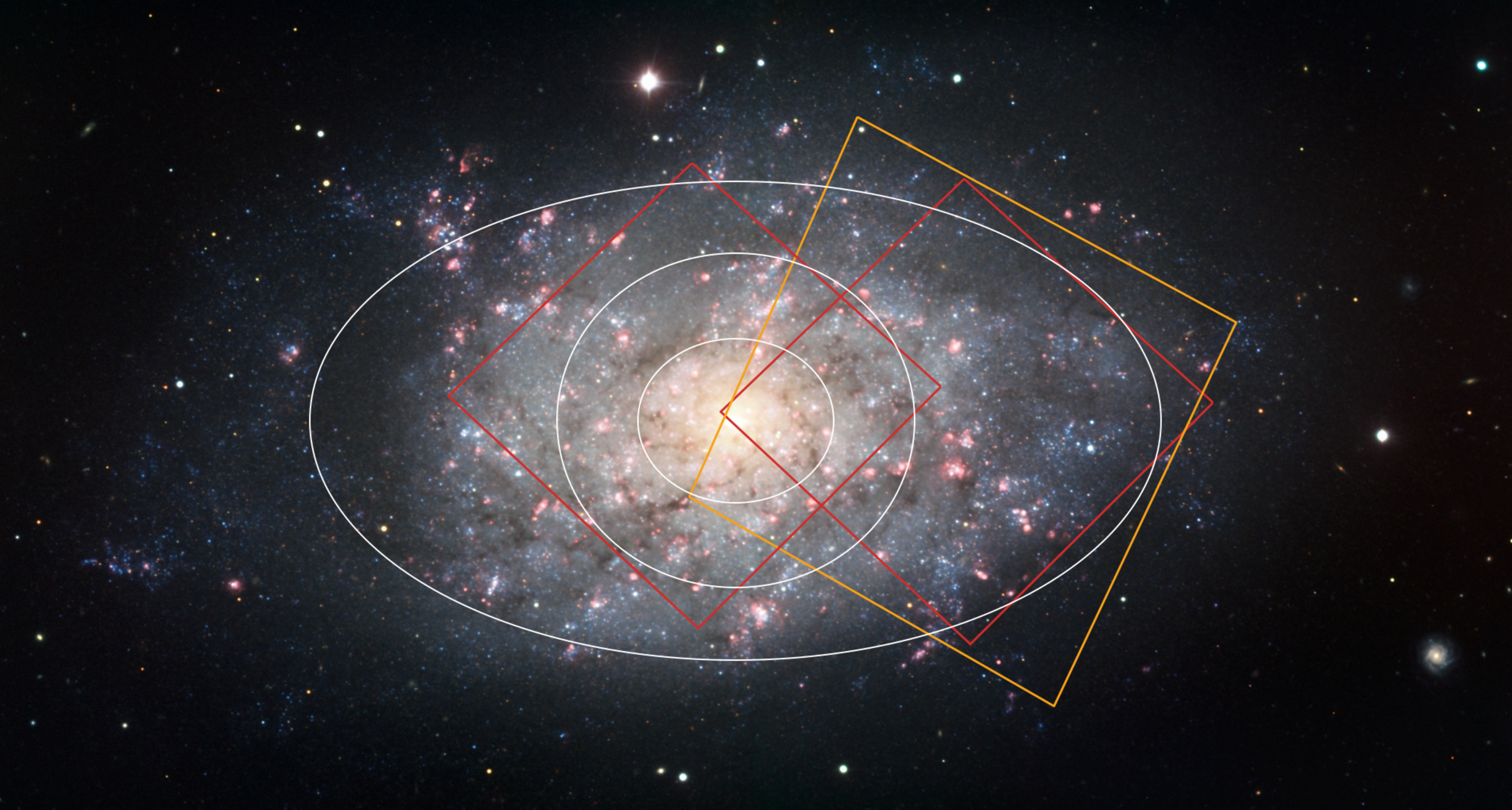}
\caption{UVIS (red) and ACS (orange) footprints overplotted on a color combined image of NGC~7793, observed with the FORS instrument at the ESO's Very Large Telescope. Overplotted in white, the 3 ellipses we used to divide the galaxy in radial regions and  recover the SFH. The image is based on data obtained through B, V, I and H$\alpha$ filters (north is up, east is left. Credit: ESO).}
\label{color_7793}
\end{figure*}

\begin{figure}
\centering
\includegraphics[width=\linewidth]{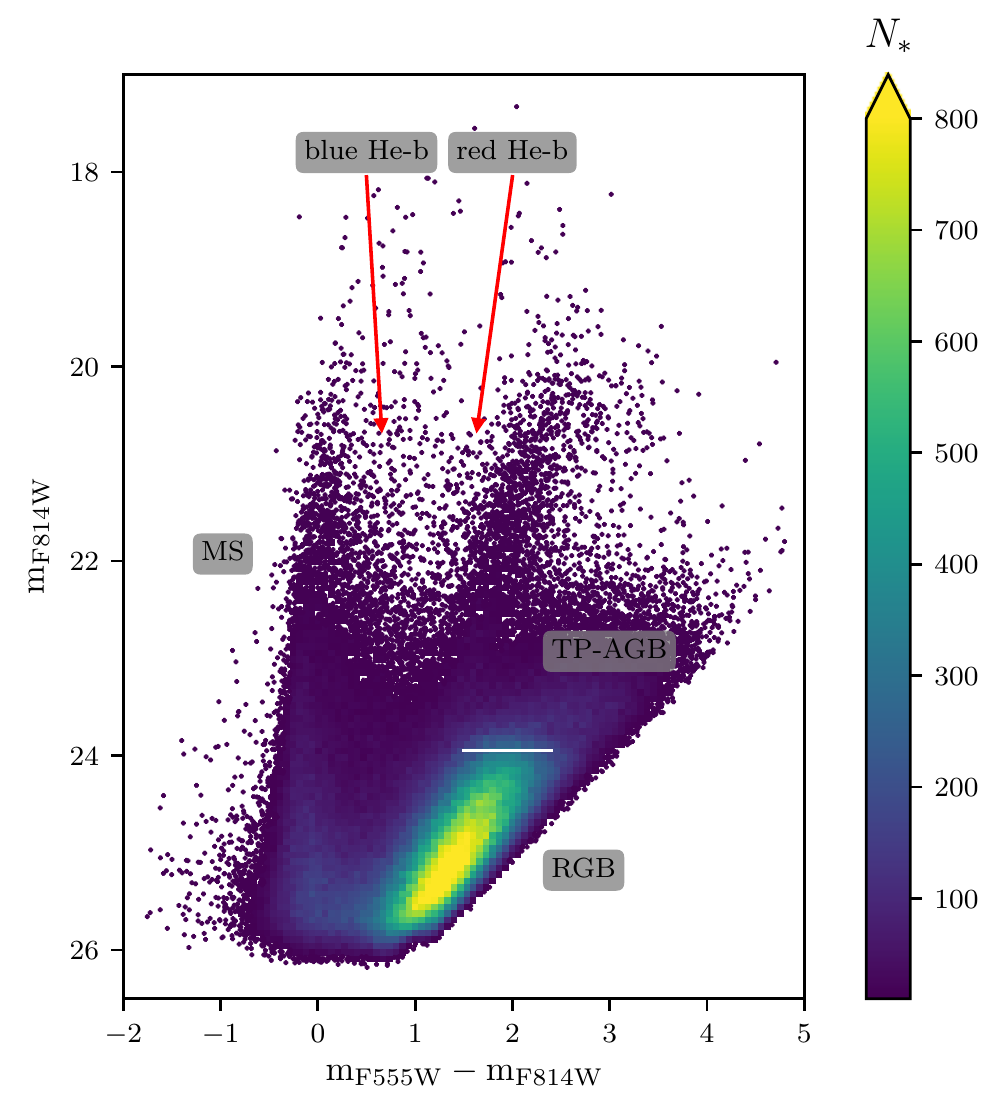}
\caption{V/I Color-Magnitude Diagram (after the quality cuts) of the two fields of NGC~7793 covered by the WFC3 and ACS imaging. The high density regions have been binned and color coded by number density (see the color-scale bar on the right of each panel) for a better visualization of the evolutionary features in the diagram. The main stellar evolutionary phases are indicated (see Section \ref{sec:7793_pop}). The horizontal white line represents the magnitude of the red giant branch tip.}
\label{7793_cmd}
\end{figure}

\begin{figure}
\centering
\includegraphics[width=\linewidth]{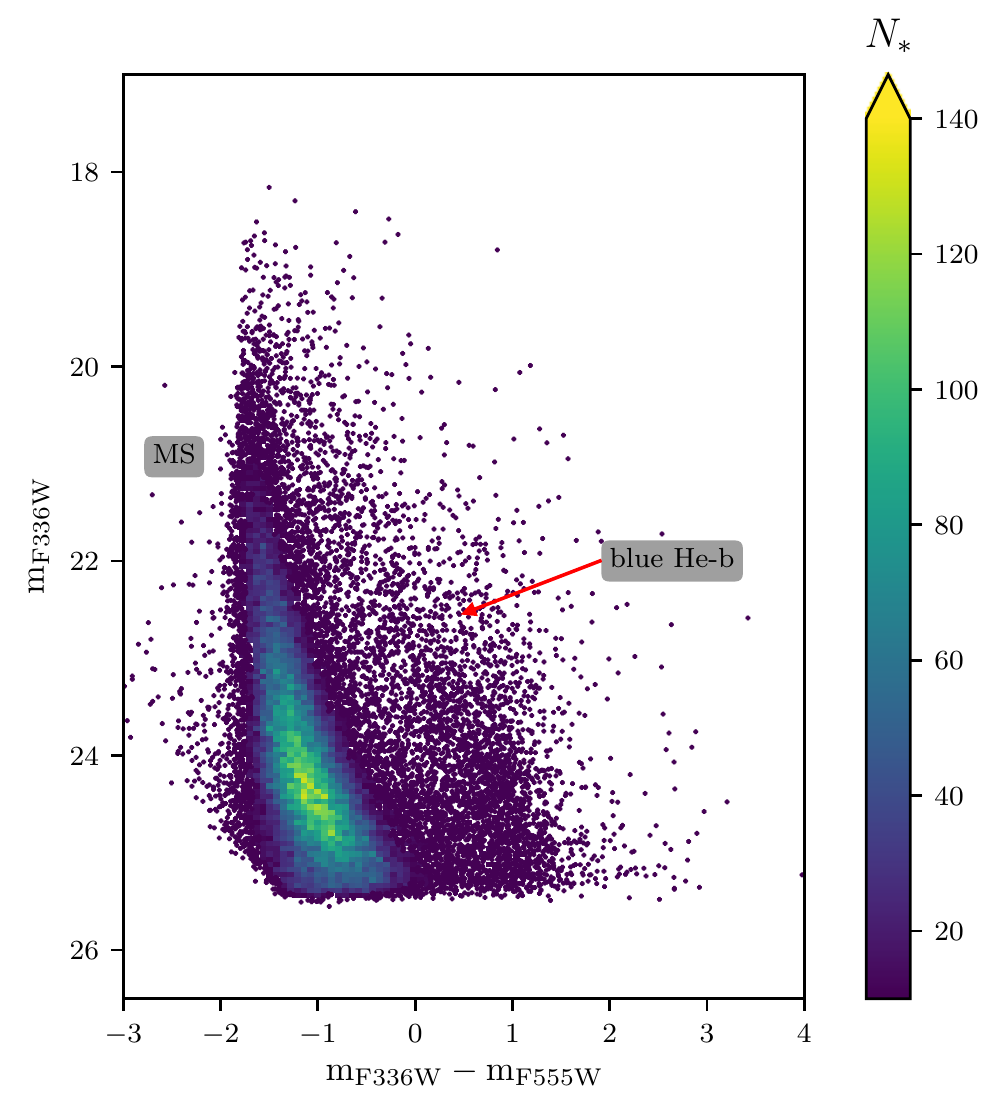}
\caption{U/V Color-Magnitude Diagram (after the quality cuts) of the two fields of NGC~7793 covered by the WFC3 imaging. The high density regions have been binned and color coded by number density as in Figure \ref{7793_cmd}.}
\vspace{1cm}
\label{7793_cmd_uv}
\end{figure}

From VLA H~\textsc{i} observations \citet{Carignan1990} derived the rotation curve out to $\sim 8 \arcmin$. A remarkable result of those observations is that, contrary to most flocculent spirals, the rotation curve is not flat in the outer parts but appears to be declining ($\Delta V_{rot} \simeq 30$ km/s or 25\% of $V_{max}$ between the maximum velocity and the last point of the rotation curve), even after modeling (tilted-ring model) a warp in the outer H~\textsc{i} disk. Though declining in the outer parts, the rotation curve is flatter than a pure Keplerian decline, and a dark halo is still needed to properly model the mass distribution. \citet{Dicaire2008} followed up this study and confirmed the uniqueness of this rotation curve with independent observations using H$\alpha$ as tracer. On the other hand, \citet{deBlok2008} found a more gentle decline of the rotation curve in the outer parts with respect to \citet{Carignan1990}, and considered it still consistent with a flat rotation curve. More recently, \citet{Bacchini2019} re-analyzed the same data and interpreted the declining rotation curve as the result of a small line-of-sight warp, finding a parametric mass model for the dark matter halo that allows them to well reproduce the whole rotation curve.

NGC~7793 has also been studied using its molecular gas emission. \citet{Muraoka2016} performed CO($J=3-2$) observations with the Atacama Submillimeter Telescope Experiment, mainly to explore correlations among its CO emission, IR luminosity, and SFR. \citet{Grasha2018} studied the CO($J=2-1$) emission traced by the Atacama Large Millimeter/submillimeter Array (ALMA) to investigate the relation between giant molecular cloud properties and the associated stellar clusters in the galaxy; they find that younger star clusters are substantially closer to molecular clouds than older star clusters, and, as expected, preferentially located on the spiral arms.\\

Here, we use both archival and new observations from the \textit{Hubble Space Telescope} (\textit{HST}) Legacy ExtraGalactic UV Survey (LEGUS; \citealt{Calzetti2015}) to study the stellar populations of this galaxy and infer for the first time its spatially resolved star formation history employing the synthetic CMD method.


\section{Observations and Data} \label{sec:7793_obs}
LEGUS observed NGC~7793 with the \textit{HST}/WFC3 UVIS channel, in the filters F275W (NUV), F336W (U), F438W (B), F555W (V), and F814W (I), in the two fields shown in red in Figure \ref{color_7793}. For the western field, archive ACS observations were already available in F555W and F814W (orange footprint in the figure). The images were aligned, drizzled, combined, and finally processed with the photometric package DOLPHOT version 2.0 \citep{Dolphin2016}. More details regarding the LEGUS photometric catalogs can be found in \citet{Sabbi2018}.

From these data, we built two independent sets of catalogs, one containing all the stars with fluxes in both F336W and F555W (U/V catalog), the other containing all those with fluxes in both F555W and F814W (V/I catalog).

\begin{figure*}
\centering
\includegraphics[width=\linewidth]{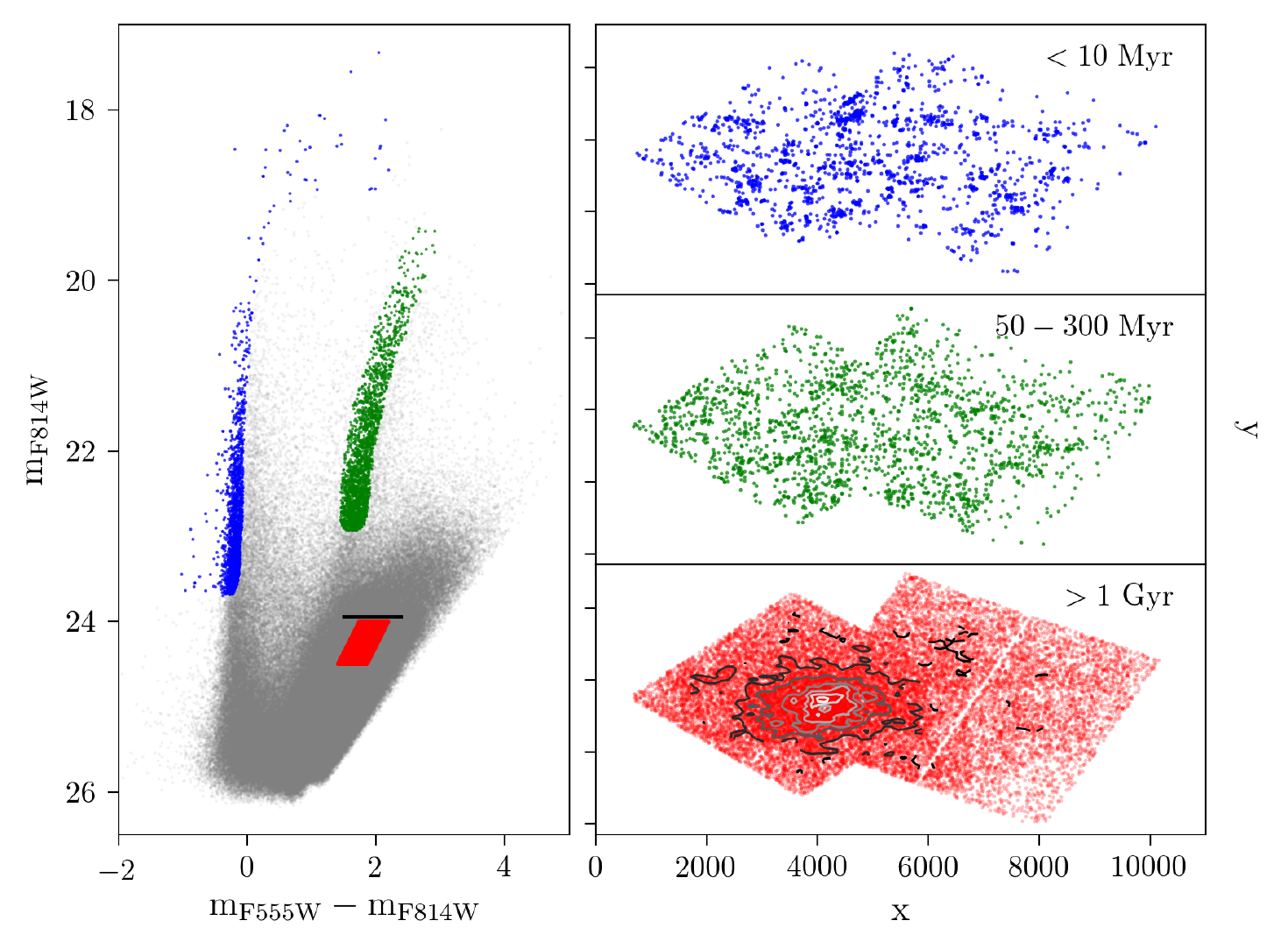}
\caption{\textit{Left panel.} Selection of three different stellar populations in the V/I CMD: in blue, stars with ages $< 10$ Myr; in green, stars with intermediate ages between $\sim 50$ and  $\sim 300$ Myr; in red, stars older than $1$ Gyr. The horizontal black line represents the magnitude of the RGB tip. \textit{Right panels.} Spatial distribution of the age-selected stars. Overplotted in gray scale on the oldest stars (bottom map), we show some isodensity contours to highlight the elliptical halo of the galaxy, and the central lower density caused by incompleteness.}
\label{7793_pop}
\end{figure*}

For each star DOLPHOT provides the position relative to the drizzled image, the magnitude and a series of diagnostics to evaluate the quality of the photometry, including signal-to-noise (S/N) ratio, photometric error, $\chi^2$ for the fit of the PSF, roundness (which can be used to identify extended objects), object type (which describes the shape of the source), and an error flag, which is larger than zero whenever there is an issue with the fitting (e.g., because of saturation or extension of the source beyond the detector field of view). To correctly interpret the properties of the stellar populations found in the galaxy, we need to differentiate as much as possible the bona fide candidate single stars from extended objects, blended sources, and spurious detections. Thus, we used these output parameters to select our data, using the following criteria, to retain a source: photometric error $\sigma \le 0.2$, squared sharpness $\le 0.2$, crowding $\le 2.25$, object type $\le 2$ for the F555W and F814W catalogs; photometric error $\sigma \le 0.2$, squared sharpness $\le 0.15$, crowding $\le 1.3$, object type $\le 2$ for the F336W and F555W catalogs \citep[following][]{Williams2014}.

Our final V/I catalog contains $\sim 328\,400$ stars in both F555W and F814W; the U/V catalog contains $\sim 52\,400$ stars in both F336W and F555W. In both cases we matched the stars detected in the overlapping areas of the two fields (see Figure \ref{color_7793}) to obtain a single catalog for the whole galaxy. In the V/I case, after the match, we chose to retain the WFC3 sources when available; the sources observed with ACS only were transformed to the WFC3 system following the instrument science reports\footnote{\url{http://www.stsci.edu/hst/acs/documents/isrs/isr1710.pdf}\\} provided by the Space Telescope Science Institute in order to create a consistent photometric data set. In both cases we checked not to have any shifts or systematic effects between the matched stars.

The CMDs corresponding to the two catalogs are shown in Figures \ref{7793_cmd} and \ref{7793_cmd_uv}.


\begin{figure*}
\centering
\includegraphics[width=\linewidth]{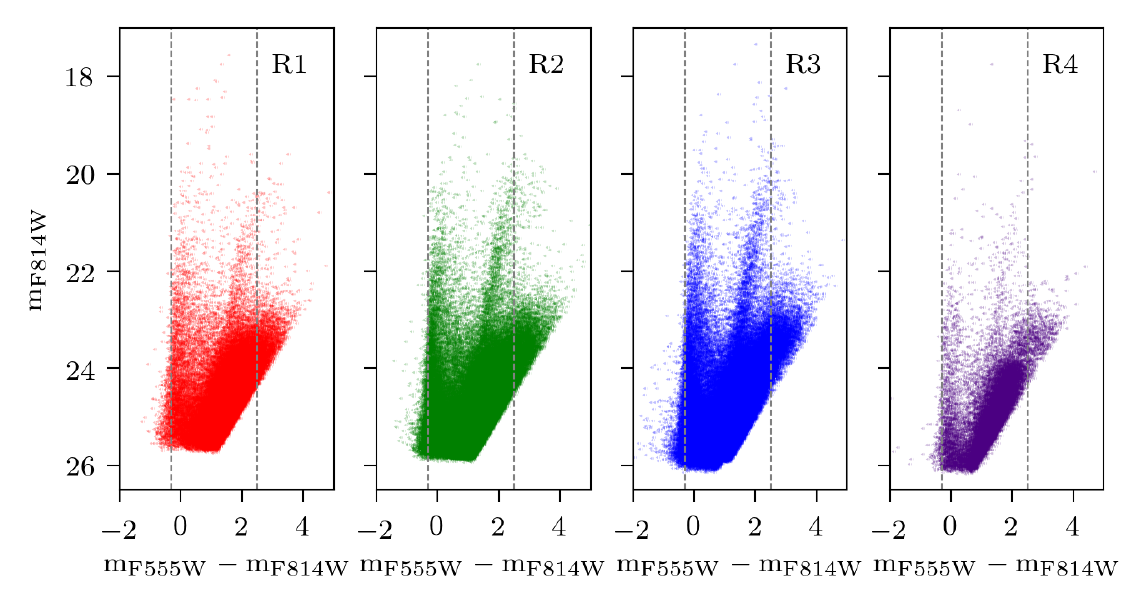}
\caption{V/I CMDs of the 4 radial regions we identified in the galaxy (see Figure \ref{color_7793}) and used to recover the SFH (R1 being the innermost ellipse, R4 the outermost annulus); the dotted vertical lines are a rough reference of the blue edge of the MS ($\mathrm{m_{F555W}-m_{F814W}}=-0.3$) and the red edge of the RGB ($\mathrm{m_{F555W}-m_{F814W}}=2.5$).}
\label{7793_fields}
\end{figure*}

\begin{figure*}
\centering
\includegraphics[width=\linewidth]{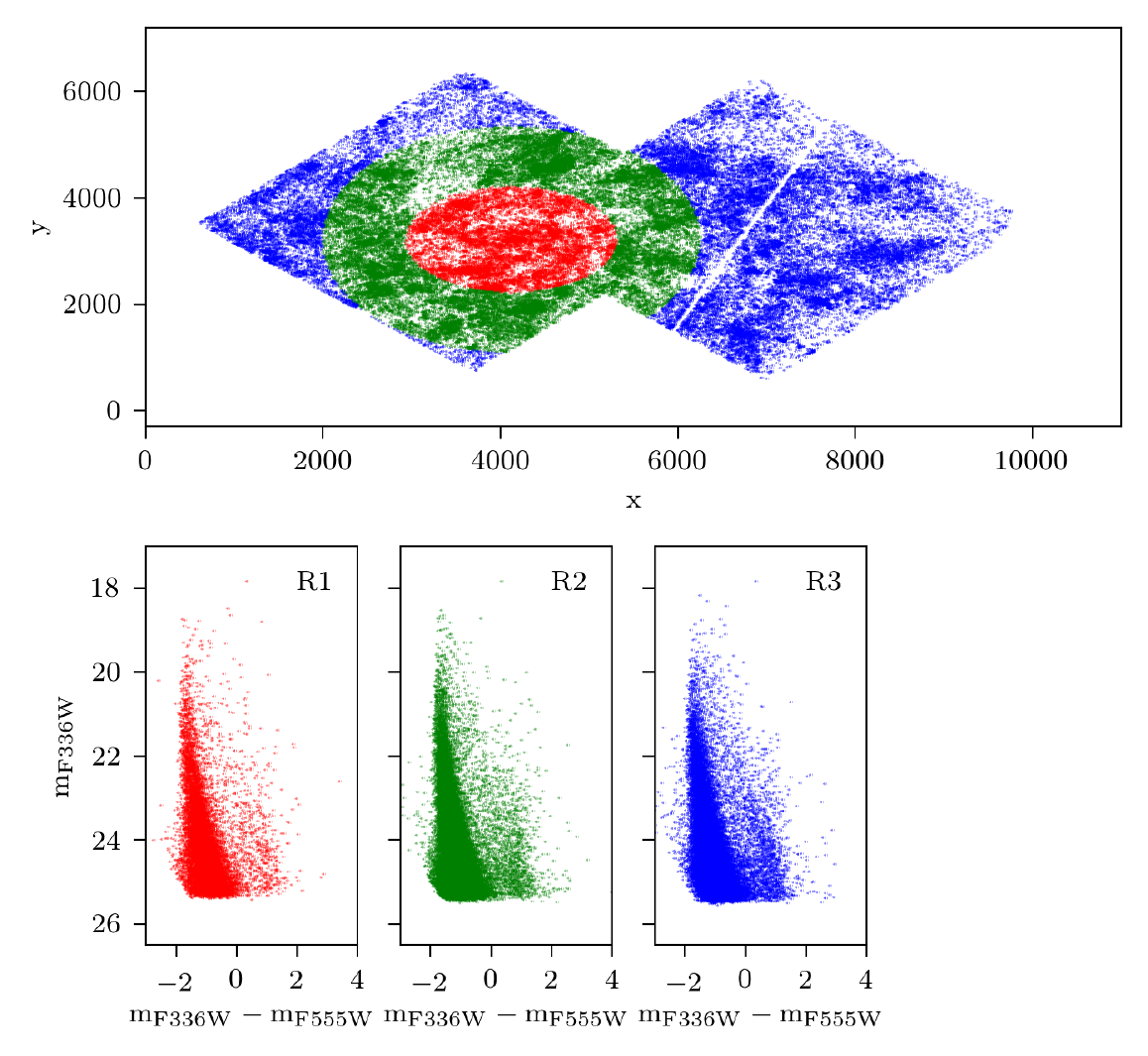}
\caption{Same regions as in Figure \ref{7793_fields} for the U/V catalog. Notice that this area was covered by two WFC3 fields, while the area in Figure \ref{7793_fields} was covered by one WFC3 field and one ACS field, hence the slightly larger spatial coverage (see also Figure \ref{color_7793}).}
\label{7793_fields_uv}
\end{figure*}

\section{Distribution of the stellar populations} \label{sec:7793_pop}
The color combined image of NGC~7793 shown in Figure \ref{color_7793} includes in red the H$\alpha$ emission, that traces the star forming H~\textsc{ii} regions of the galaxy. The figure highlights very well the flocculent morphology of the galaxy, characterized by very clumpy emission and no obvious spiral arms.

Figure \ref{7793_cmd} shows the F814W versus F555W$-$F814W CMD of the galaxy. The main stellar evolutionary phases are clearly recognizable in the CMD, and are typical of galaxies with a rather continuous star formation, from ancient to recent epochs. We see a well-populated blue plume, with main sequence (MS) and hot core He-burning stars, the red plume with mainly red He-burning stars, some blue loop (BL) stars between the two plumes, the horizontal feature characteristics of thermally pulsing asymptotic giant branch (TP-AGB) stars, and finally the red giant branch (RGB), our older age signature. The magnitude of the RGB tip is also indicated with a horizontal segment.

As a first approach to study how different stellar populations are distributed within the galaxy, we isolated three age intervals in the CMD and plotted the selected stars on the spatial map. The results are shown in Figure \ref{7793_pop}.

As expected for a spiral galaxy, the young (age $<10$~Myr, following a PARSEC isochrone \citep{Bressan2012,Marigo2017} to select the brightest and bluest part of the blue plume) stars follow the structure of the flocculent spiral ``arms'', spread all over the disk of the galaxy; the majority ($\sim 93$\%) of the sources found in this age interval also have a measured flux in the F336W filter. The slightly older (ages between $50$ and $300$~Myr, from the red plume selection) population shows a similar, but less clumpy distribution. The oldest (age $>1$~Gyr, from the RGB) stars present in our sample reveal an elliptical shape that traces the whole body of the galaxy, with a central hole due to the incompleteness of the most crowded region. It is particularly interesting that the this old star distribution do not show any spiral structures, indicating there are probably no spiral density waves, or possibly only very weak (imperceptible) waves. Thus, the spiral arms seen in visible or UV light are just star formation regions, not spiral waves. This was already pointed out by \citet{Elmegreen1984} from blue and near-infrared surface photometry, suggesting that the SF in extreme flocculent galaxies is from pure gas processes. Also, since the SF regions are not in shells, are fairly large, and show a hierarchical structure, they are probably the result of gravitational instabilities in the gas \citep{Elmegreen1990,Elmegreen1992}.

To study the spatial variations of the SF across the body of the galaxy, we divide it in four elliptical sub-regions as shown in Figure \ref{7793_fields}, following the distribution of the star forming areas in the galaxy. The same selection was applied to our U/V catalog. In this case, the outermost region (R4) contains a very low number of stars, given the smaller area covered by the two WFC3 fields (see Figure \ref{color_7793}), thus we included those stars in Region 3, as shown in Figure \ref{7793_fields_uv}.

The F336W versus F336W$-$F555W CMD (shown in Figure \ref{7793_cmd_uv}) is populated by only two main stellar phases, given the shorter lookback time reachable with these bands (a few hundreds of Myr). The MS is much brighter in this case, and represents the most populated phase we see in the U/V CMD, together with a less dense clump of hot He-burning stars that here are well separated from the MS, while in the V/I CMD the two phases are on top of each other. This is the main advantage of using the F336W filter, that also allows for a better time resolution in the SFH at the youngest epochs.

A look at the spatial distributions presented in Figure \ref{7793_fields_uv} also reveals that the U/V band combination traces a very patchy population, both because of the shorter lookback time sampled here and the effect of dust in the most active SF regions of the galaxy
(see, e.g., the work by \citealt{Kahre2018} on five LEGUS galaxies including NGC~7793).

\begin{figure}
\centering
\includegraphics[width=\linewidth]{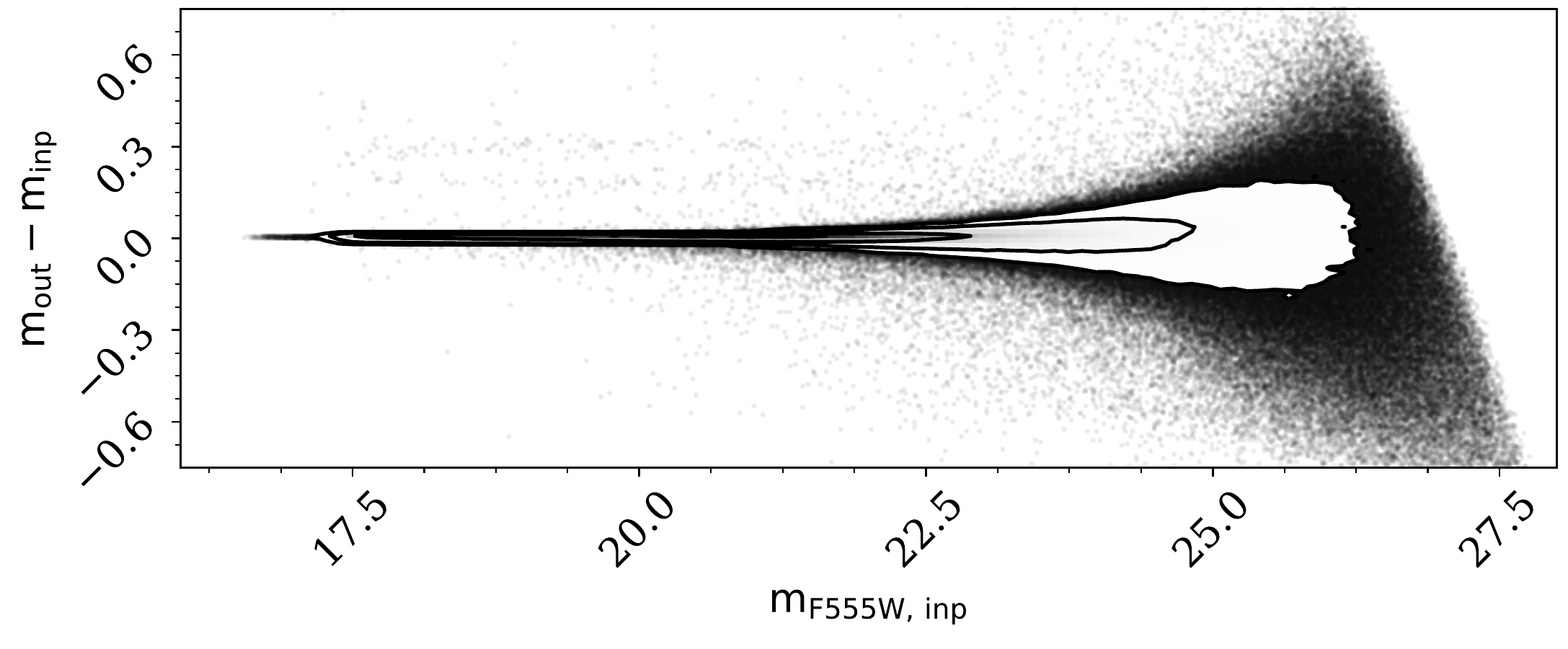}
\centering
\includegraphics[width=\linewidth]{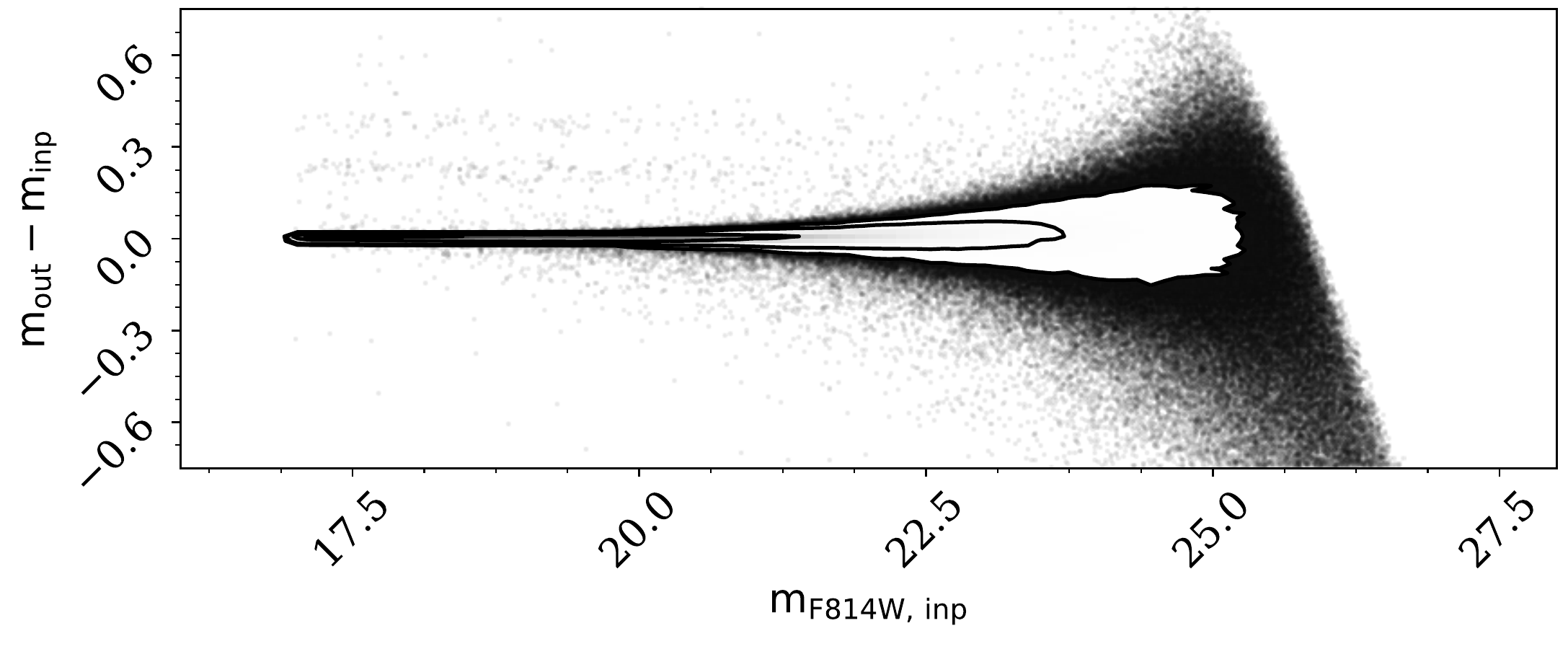}
\caption{Photometric errors in F555W (top panel) and F814W (bottom panel) from our artificial star tests; the contours indicate the 1$\sigma$, 2$\sigma$ and 3$\sigma$ levels of the distributions.}
\label{7793_ast}
\end{figure}

\begin{figure}
\centering
\includegraphics[width=\linewidth]{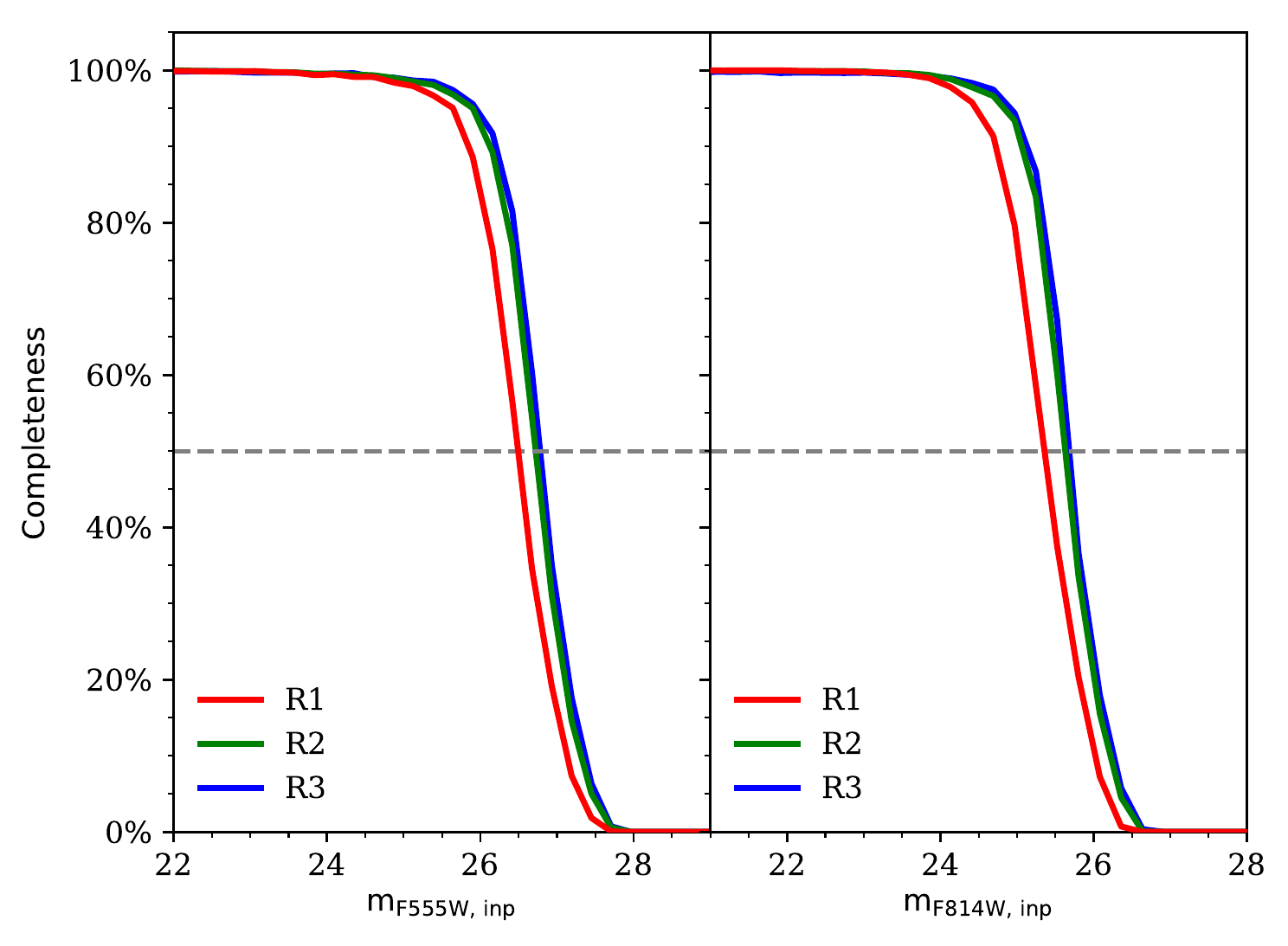}
\caption{Example of the eastern field completeness in F555W (left panel) and F814W (right panel) from our artificial star tests in the different regions of the galaxy highlighting the slightly different crowding conditions from inside out. The dashed horizontal line marks the 50\% completeness level. Notice that R4 is mostly in the western field, thus it is not represented here (its completeness, however, is very similar to that of R3).}
\label{7793_compl}
\end{figure}


\section{Artificial Star Tests} \label{sec:7793_ast}
We performed artificial star tests on the images in order to properly estimate the photometric errors and incompleteness affecting our catalogs. As can be evaluated from Figures \ref{7793_pop}$-$\ref{7793_fields_uv}, the two catalogs are quite different in terms of homogeneity of the stellar populations and density gradients within the fields. For this reason, we performed a slightly different procedure in the two regimes.

In both cases we first add to the real images one artificial star, for which we know exactly position and input magnitude, and re-run DOLPHOT. We then check whether the new source is detected or not, and its output magnitude. We iterate this process for 1.5 million times, varying the position and magnitude of the input fake stars, to explore the whole image and the full range of magnitudes we are interested in. Adding one star at a time guarantees that we are not introducing artificial crowding in the images. The output distribution of recovered stars gives us an estimate of the photometric error (from the $m_{output}-m_{input}$ versus $m_{input}$ distribution) and completeness (from the ratio between the number of output and input stars) as a function of both space and magnitude (see Figures \ref{7793_ast} and \ref{7793_compl} as an example in F555W and F814W).

In the F336W images (thus, in the F336W versus F336W$-$F555W CMD) we are sampling young ($\lesssim 2-300$~Myr) stars only, that tend to have a more clumpy distribution and are affected by a more severe crowding, thus incompleteness, than the average field star. To take this inhomogeneous distribution into account, we follow the steps first outlined in \citet{Cignoni2016} and then used in \citet{Cignoni2018} and \citet{Sacchi2018}. Let us call $C(x,y,m)$ the completeness as a function of the spatial coordinates and magnitude, $N_{obs}(x,y,m)$ the number of detected stars and $N_{true}(x,y,m)$ the real number of stars, without the effect of incompleteness. With the first run of artificial stars we can estimate a ``correction'' factor $1/C(x,y,m)$ and then use it to reconstruct the real density profile of the galaxy assuming that:
\begin{equation}
\frac{1}{C(x,y,m)} \times N_{obs}(x,y,m) \simeq N_{true}(x,y,m) \, .
\end{equation}

We then use this profile as the input of a second run of artificial stars with more stars added in the denser regions, and we repeat the artificial star test from the beginning. This new incompleteness is weighted with the density profile of the real stars, so it provides a more accurate estimate of the actual incompleteness suffered by the young stars. Notice that this is an iterative process, since the  completeness $C(x,y,m)$ from the first run of artificial stars is not the final one; the approximation improves with each iteration.

This method is similar to the procedure adopted by \citet{Johnson2016},  who follow the surface brightness of the real stars to choose the input positions for the artificial stars.

Figure \ref{7793_compl} shows the completeness as a function of magnitude for the tests we performed in F555W and F814W for the sub-regions in which we divided the V/I catalog (only the results for the eastern field are shown). The different crowding conditions from inside out are reflected in the slightly different completeness we obtain as an output of these tests; indeed, the most internal region is the most crowded thus incomplete one, while the completeness increases as we move outwards.


\begin{figure*}
\centering
\includegraphics[width=0.49\linewidth]{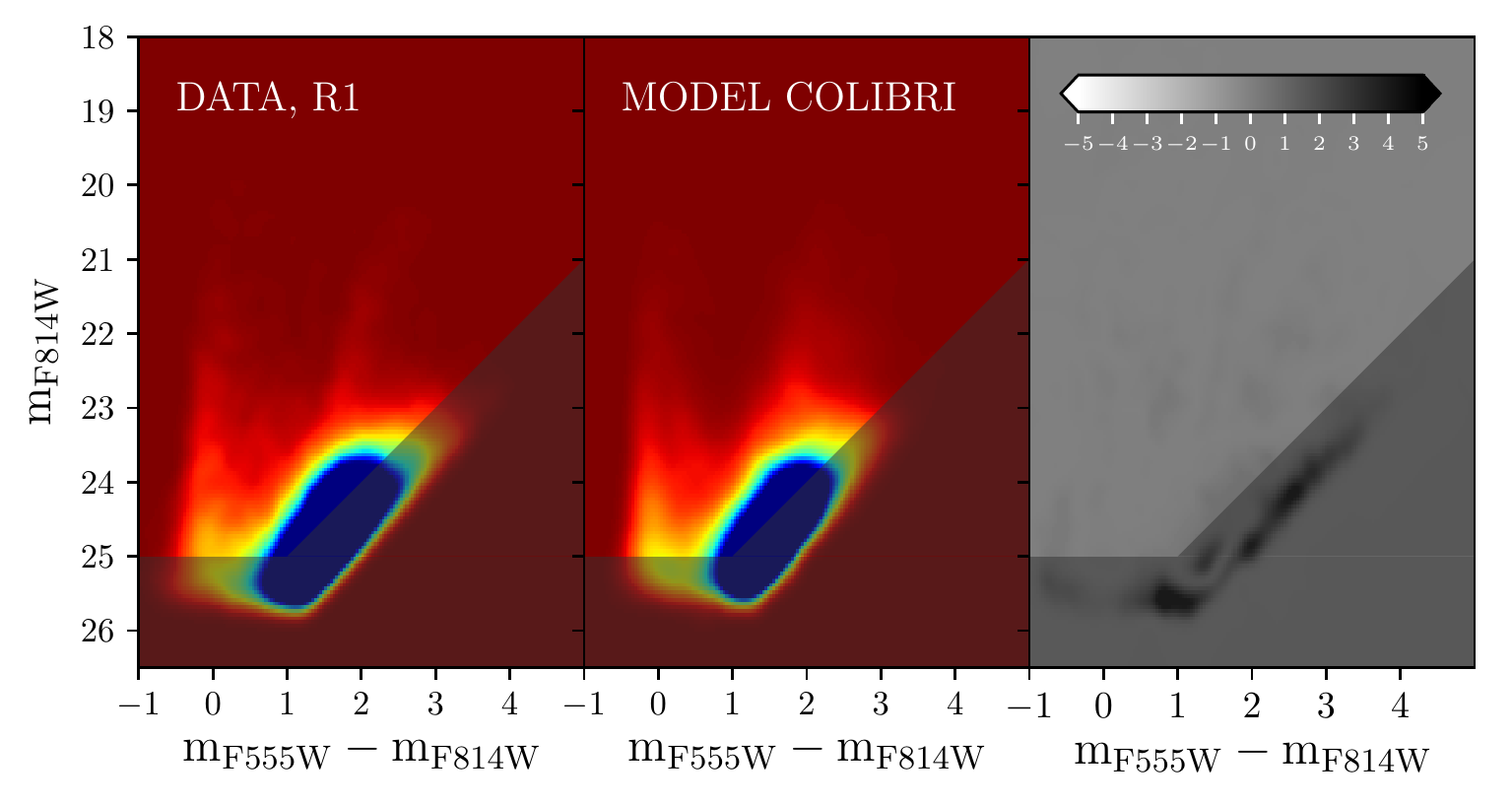}
\includegraphics[width=0.49\linewidth]{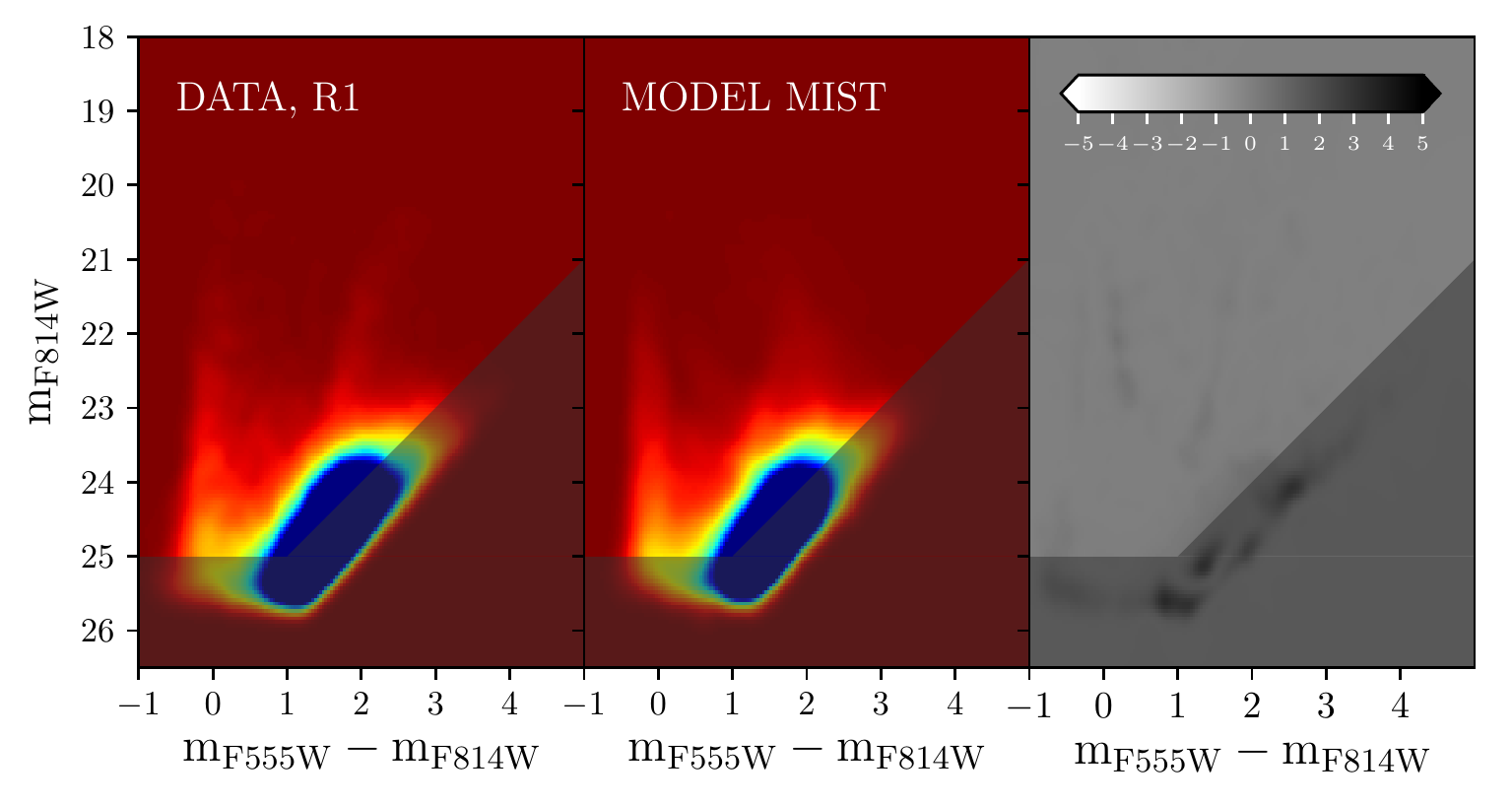}
\centering
\includegraphics[width=0.49\linewidth]{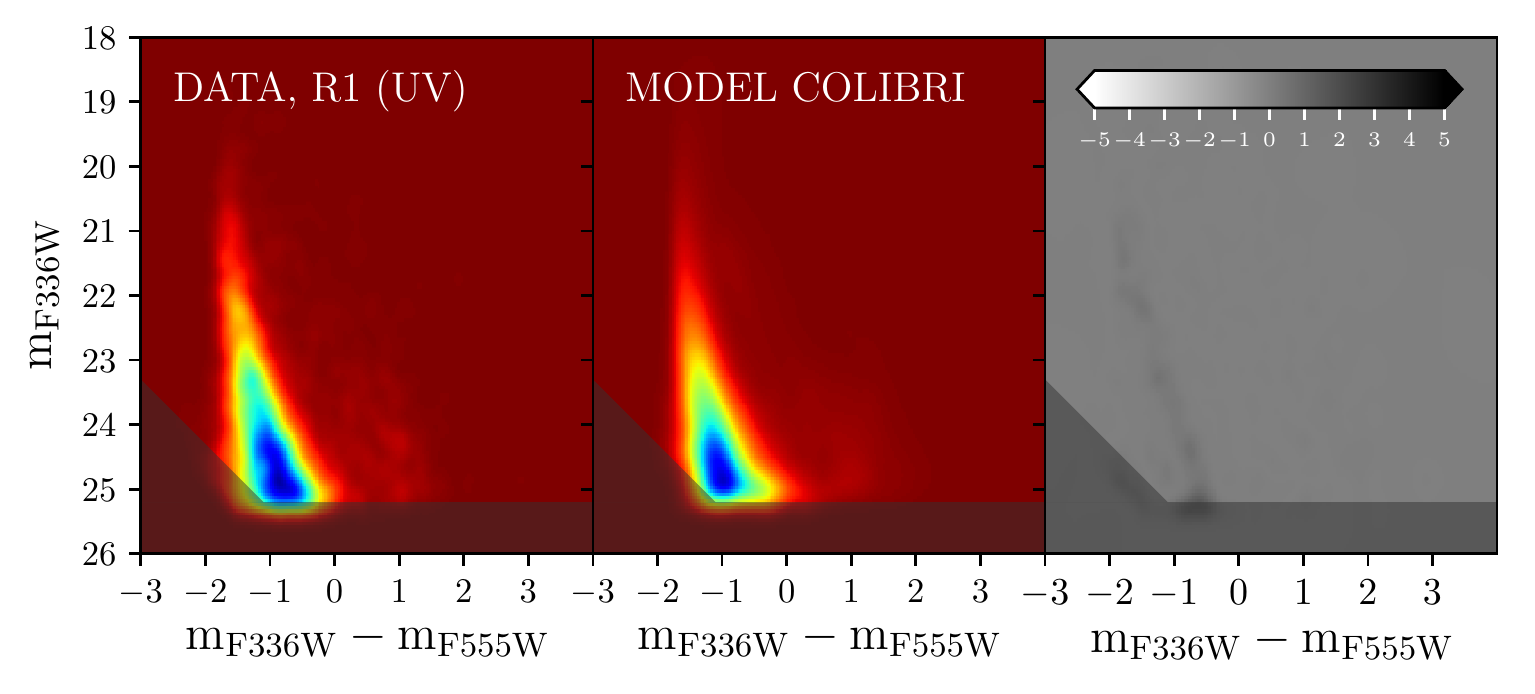}
\includegraphics[width=0.49\linewidth]{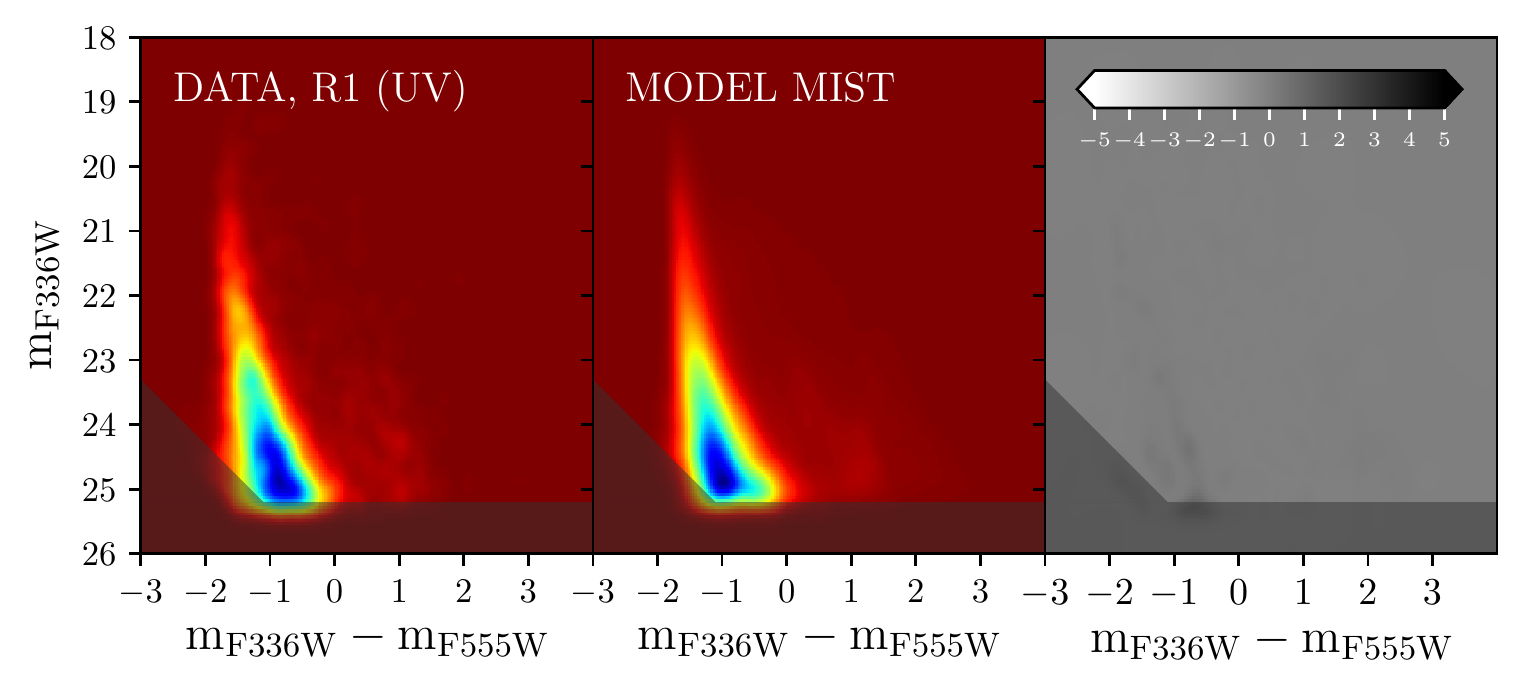}
\centering
\includegraphics[width=0.49\linewidth]{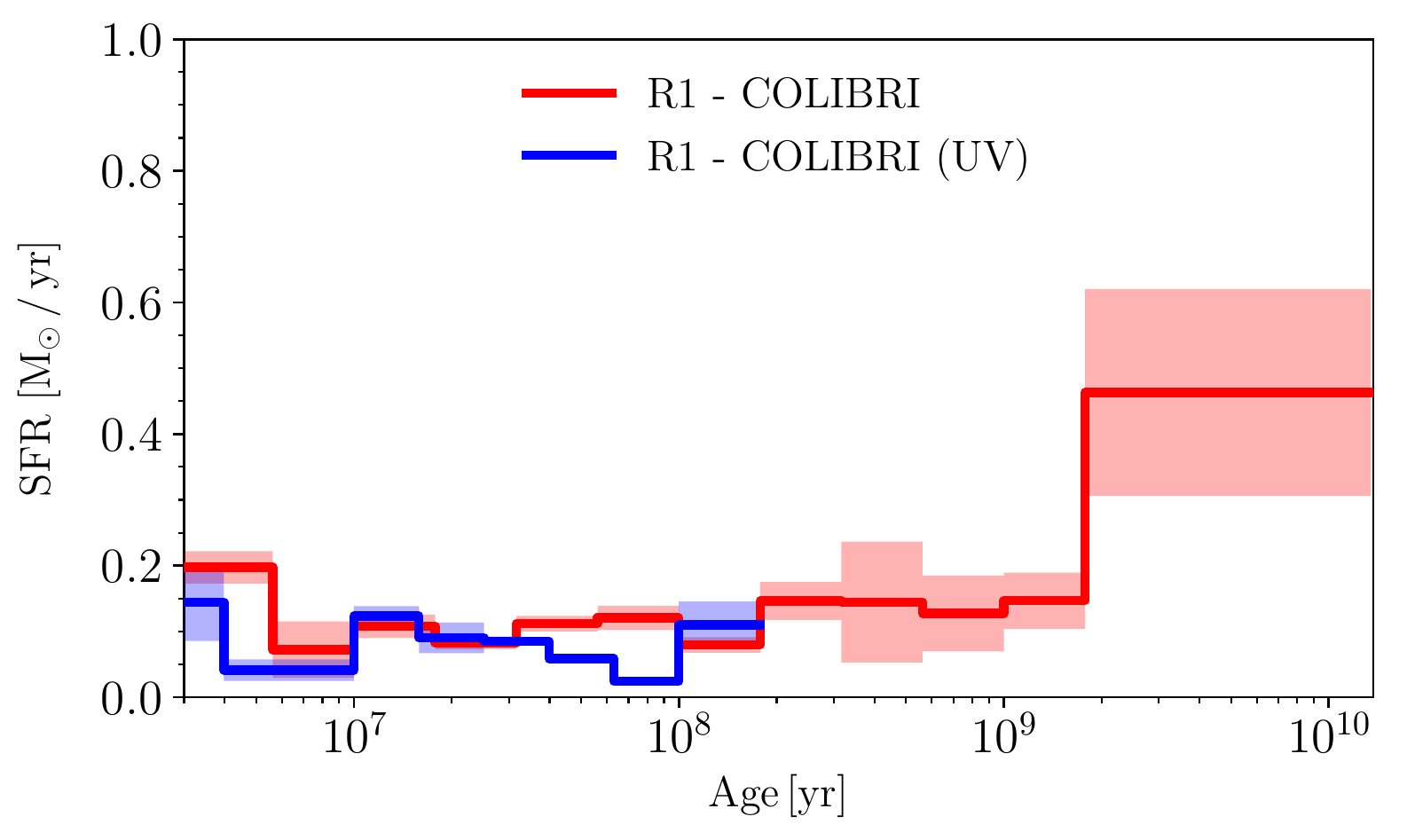}
\includegraphics[width=0.49\linewidth]{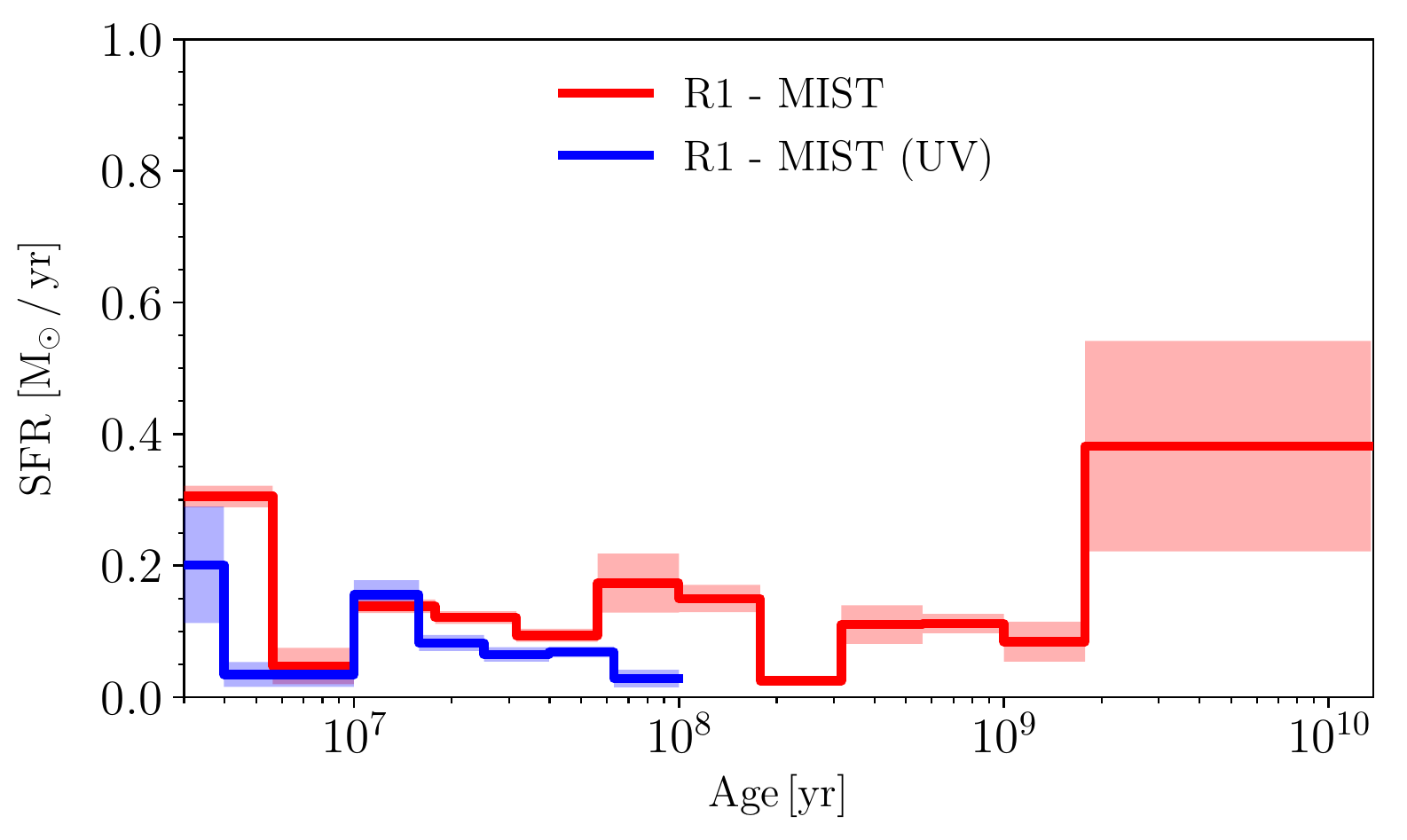}
\caption{\textit{Top panels.} Hess diagrams of the inner region of NGC~7793 from the V/I data: the observational one is on the left and the one reconstructed on the basis of different sets of models in the middle (COLIBRI models in the left-middle panels, and MIST models in the right-middle panel), while on the right we show the residuals between the two in terms of the likelihood used to compare data and models in SFERA, i.e., $data \times \ln(data/model) - data + model$; the shaded part corresponds to the area below the 50\% completeness limit used as a boundary for the SFH recovery. \textit{Middle panels.} Hess diagrams and residuals between the U/V observational and synthetic CMDs. \textit{Bottom panels.} Recovered V/I SFH in red, U/V SFH in blue, from the two sets of models (COLIBRI on the left, MIST on the  right).}
\label{7793_R1}
\end{figure*}

\begin{figure*}
\centering
\includegraphics[width=0.49\linewidth]{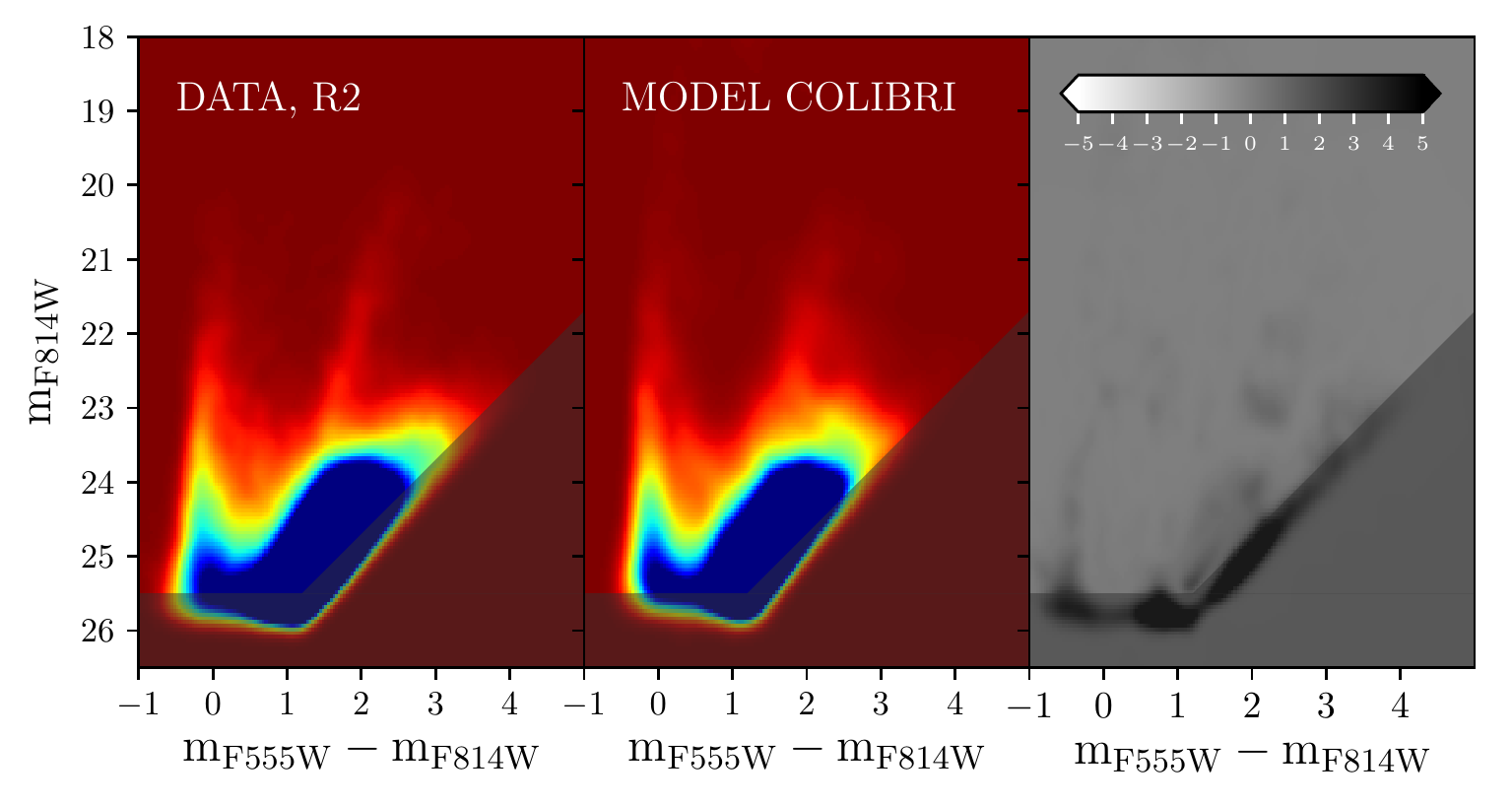}
\includegraphics[width=0.49\linewidth]{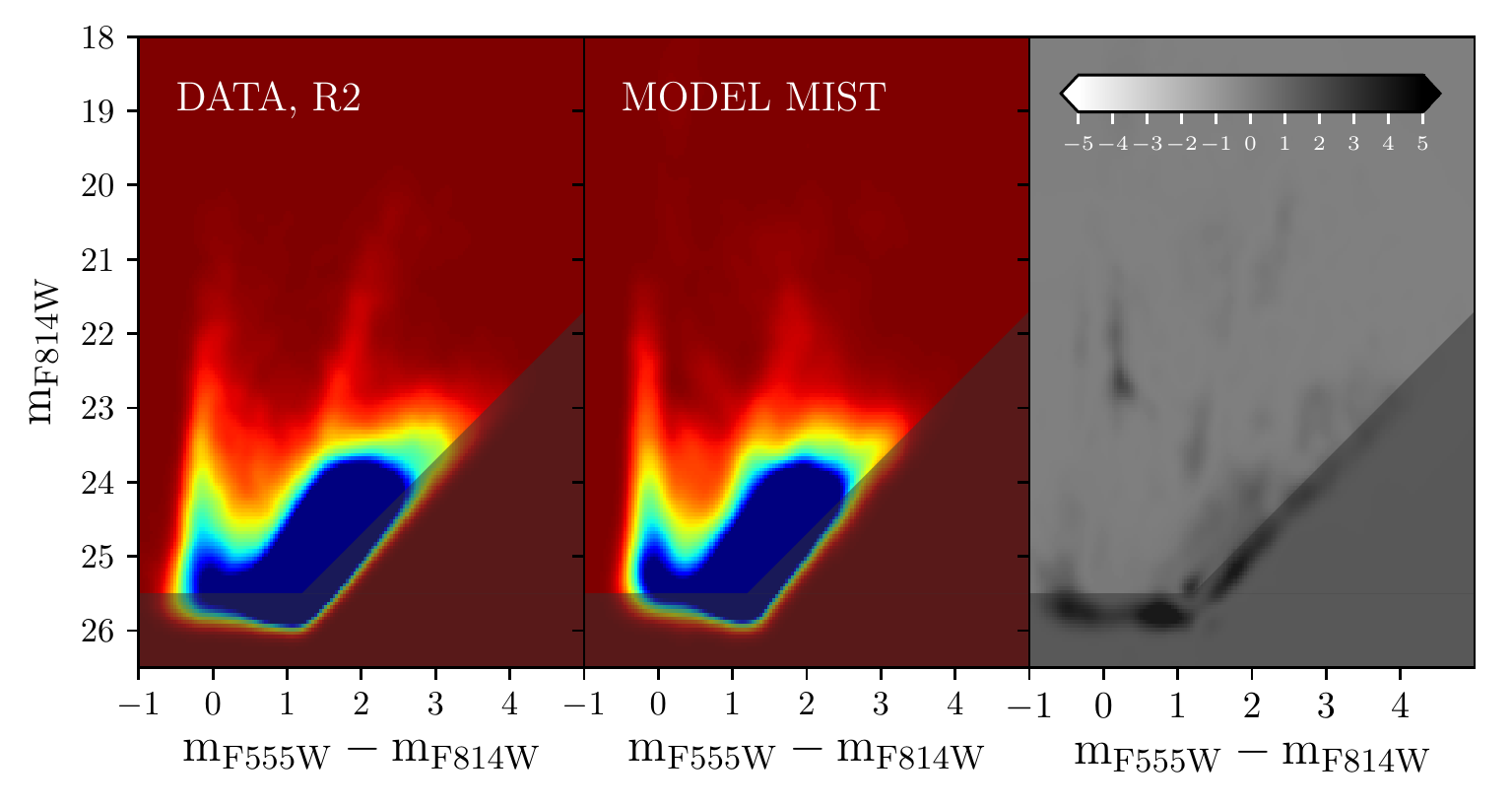}
\centering
\includegraphics[width=0.49\linewidth]{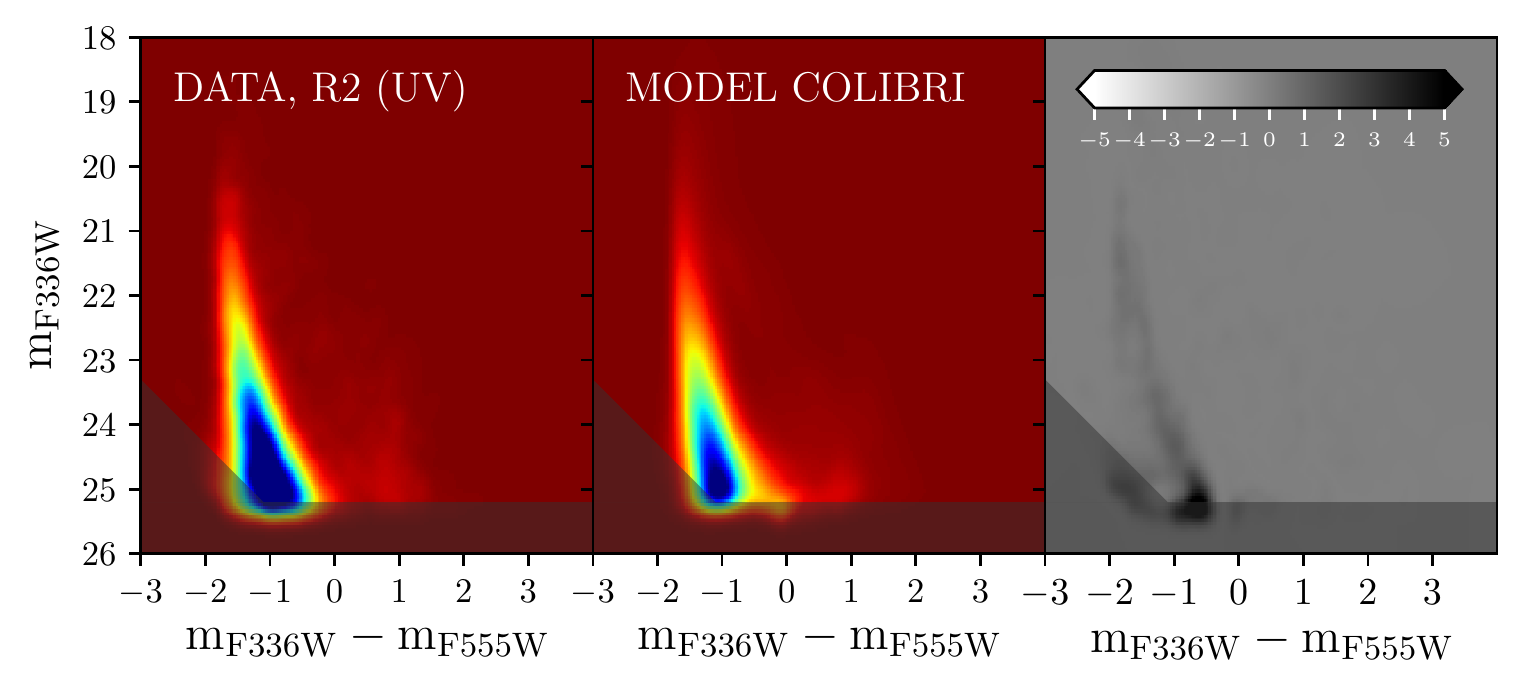}
\includegraphics[width=0.49\linewidth]{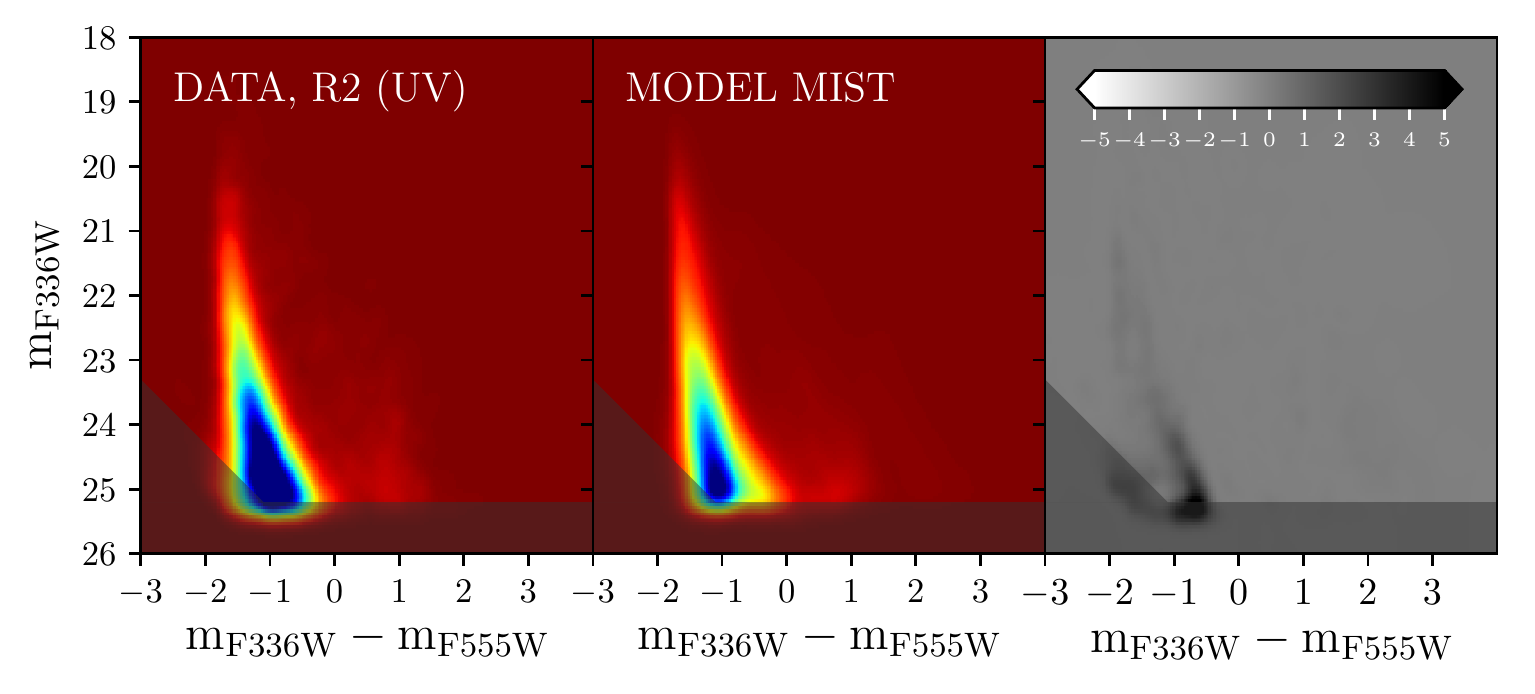}
\centering
\includegraphics[width=0.49\linewidth]{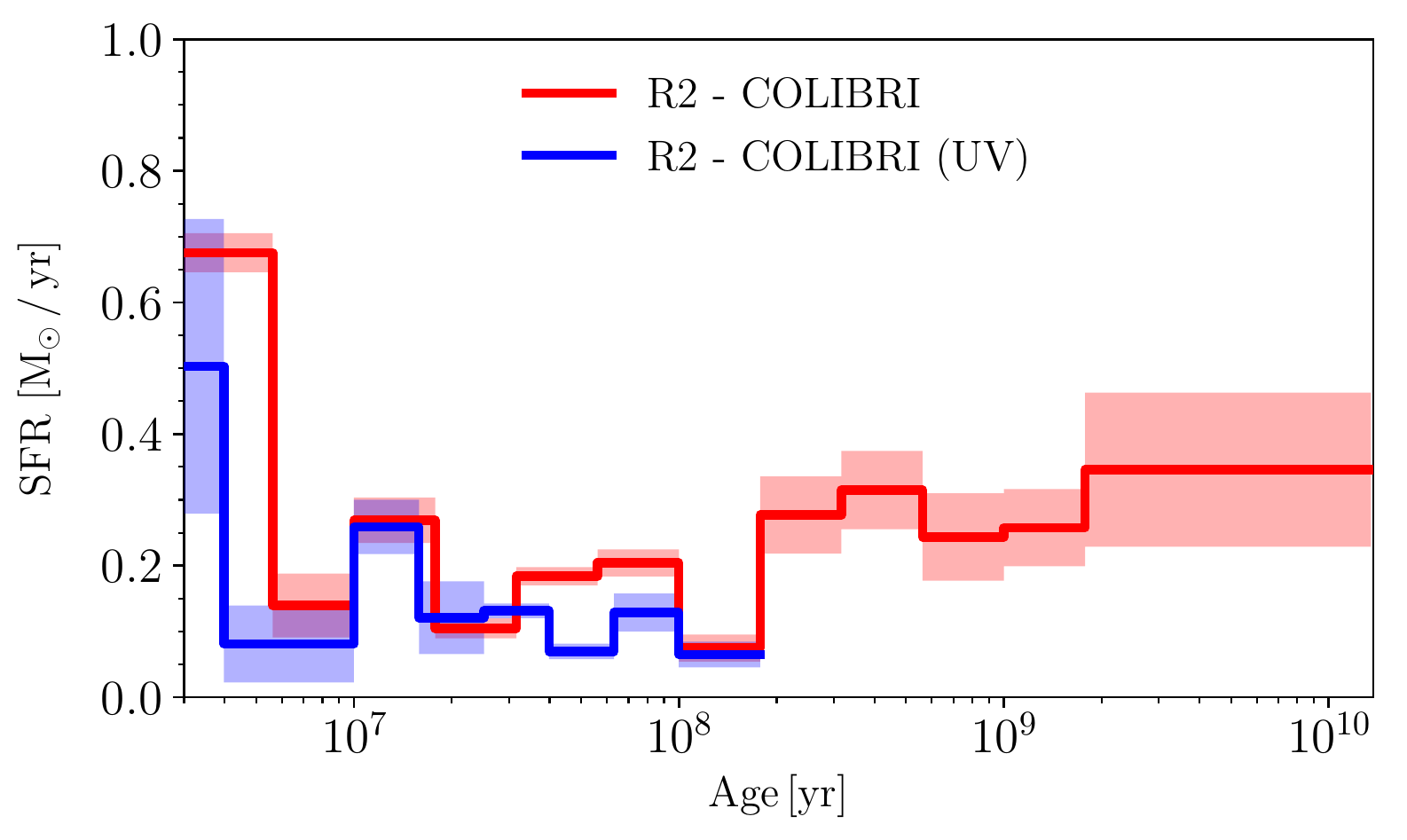}
\includegraphics[width=0.49\linewidth]{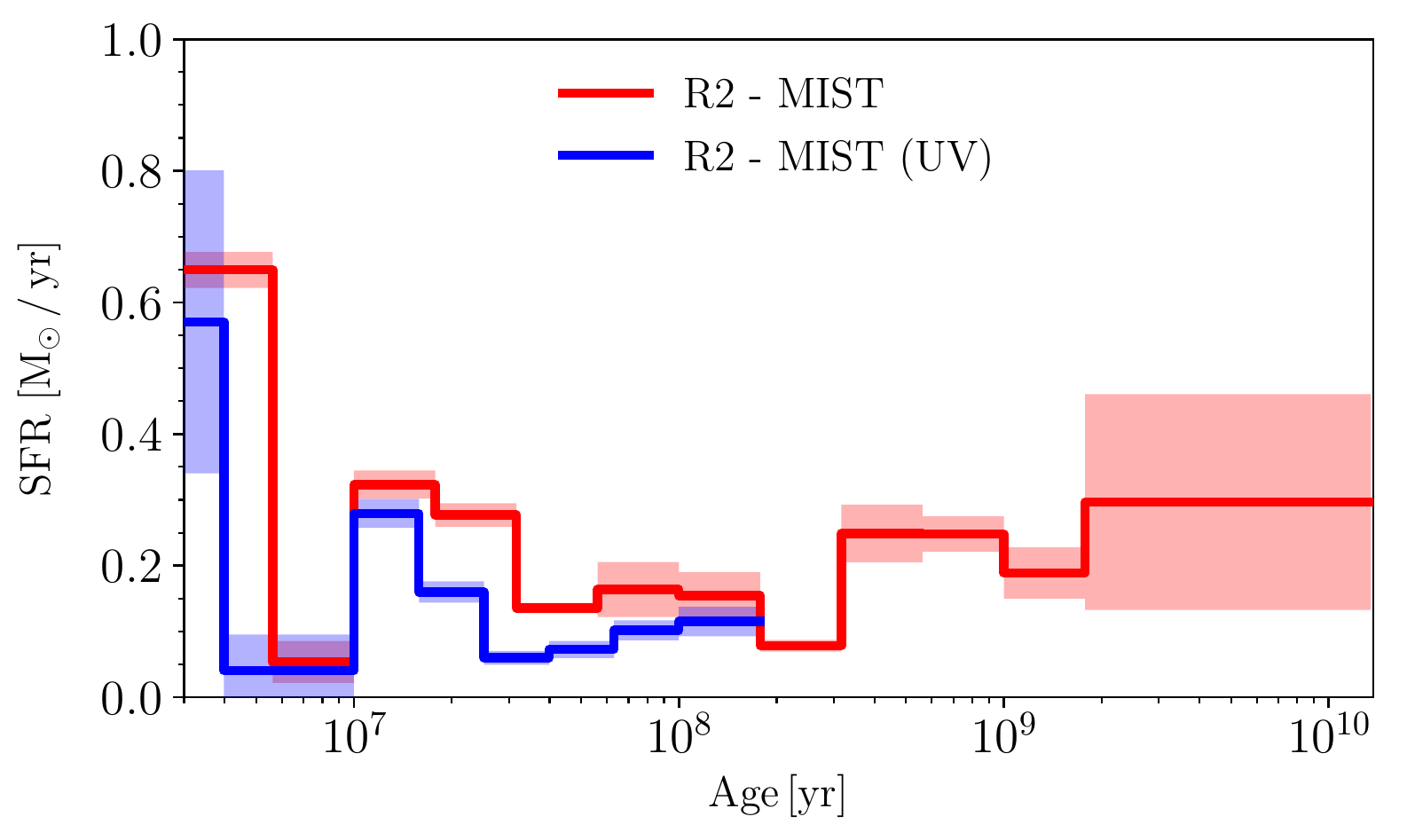}
\caption{Same as Figure \ref{7793_R1} but for the second annulus of NGC~7793.}
\label{7793_R2}
\end{figure*}

\begin{figure*}
\centering
\includegraphics[width=0.49\linewidth]{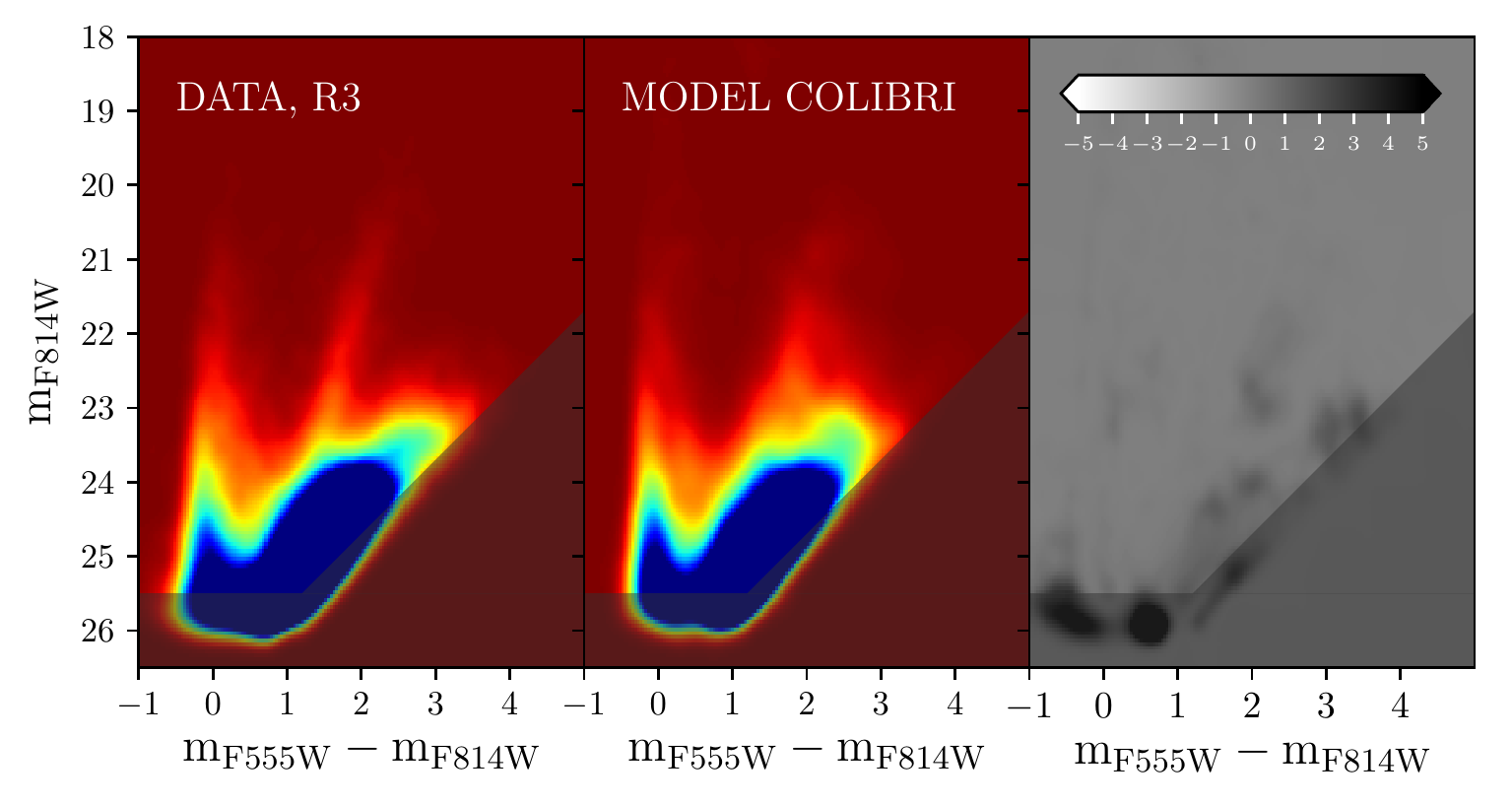}
\includegraphics[width=0.49\linewidth]{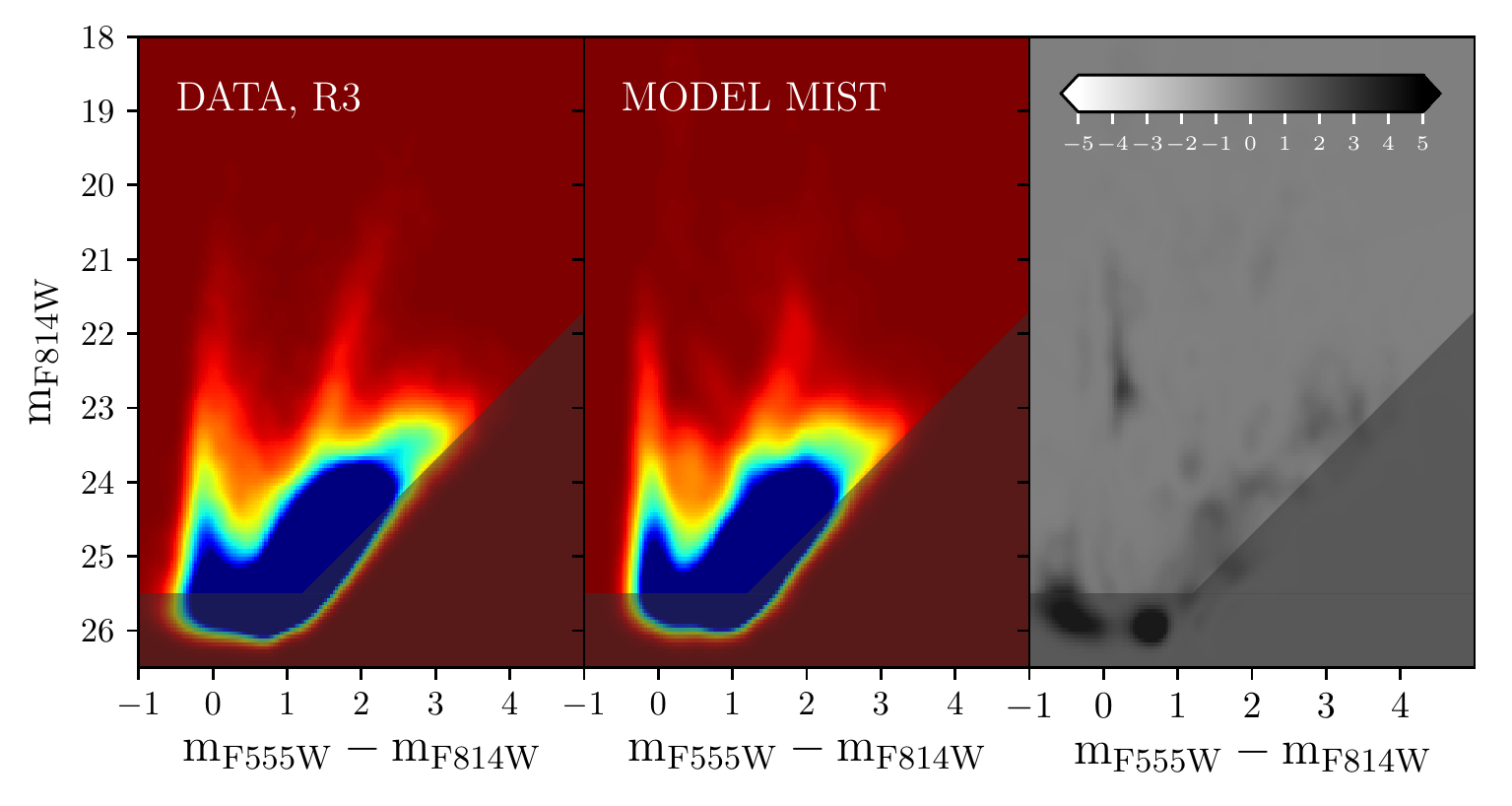}
\centering
\includegraphics[width=0.49\linewidth]{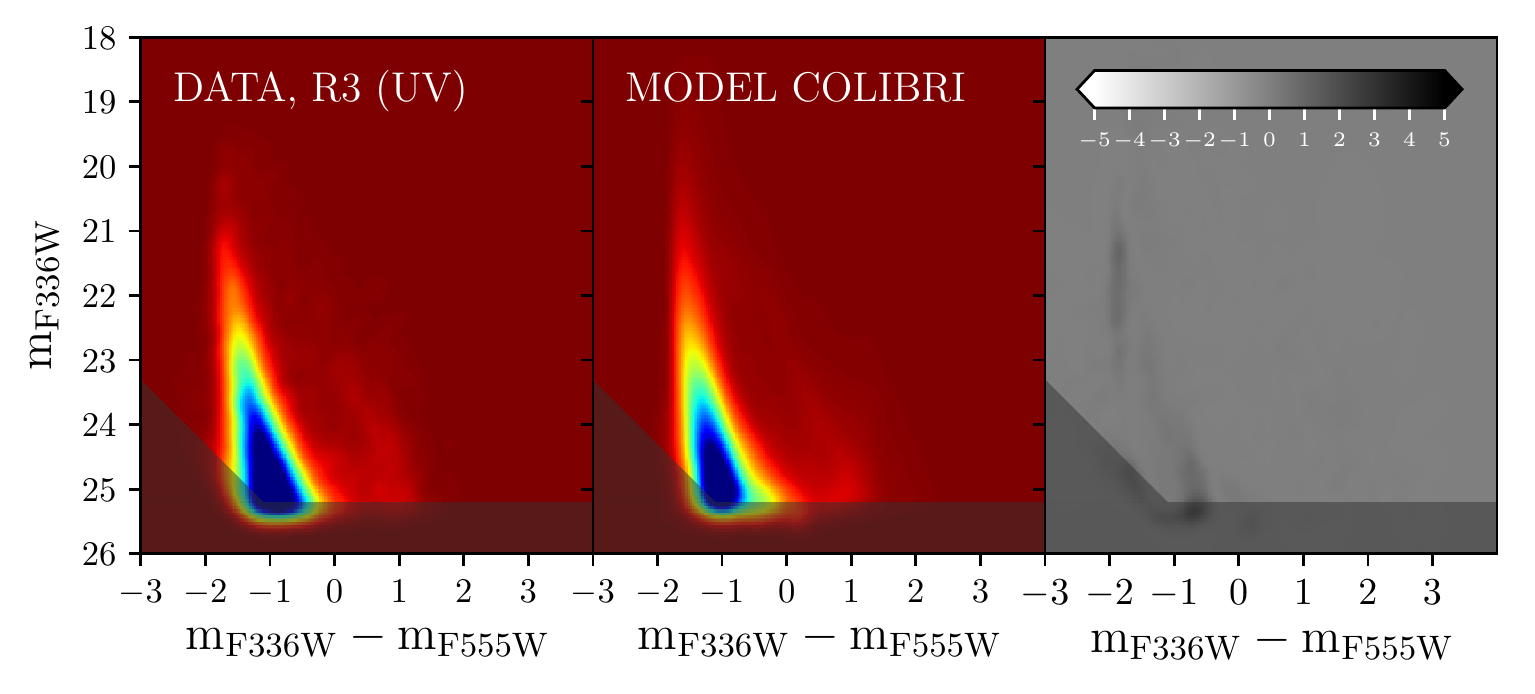}
\includegraphics[width=0.49\linewidth]{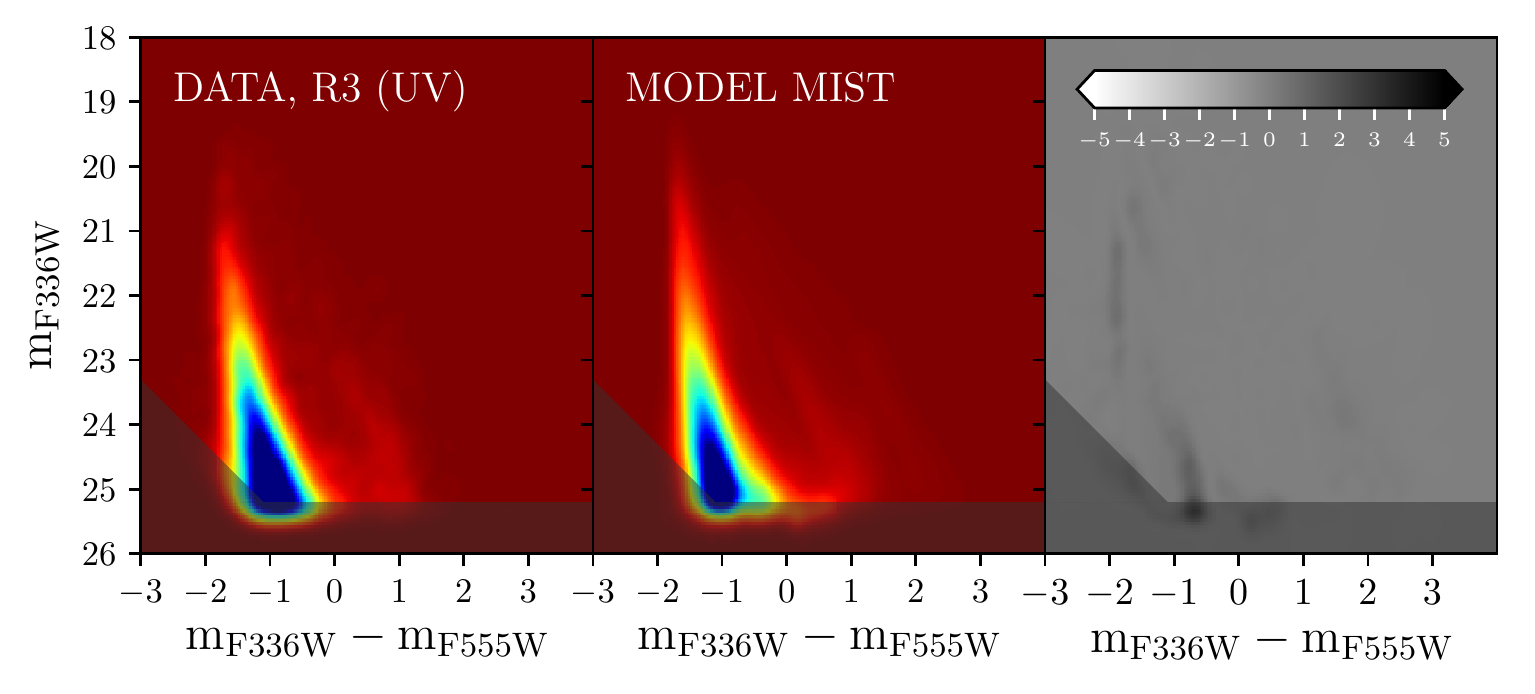}
\centering
\includegraphics[width=0.49\linewidth]{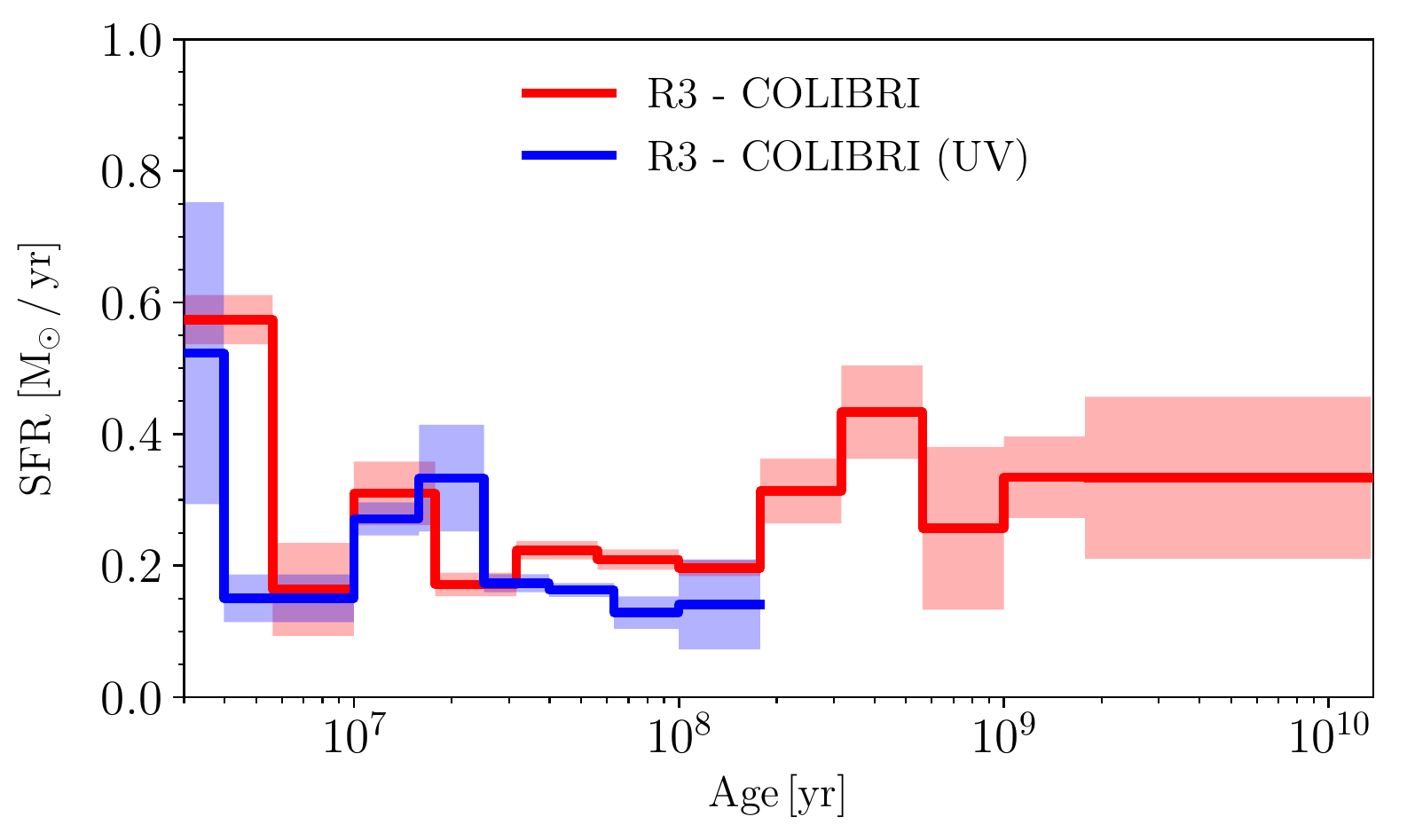}
\includegraphics[width=0.49\linewidth]{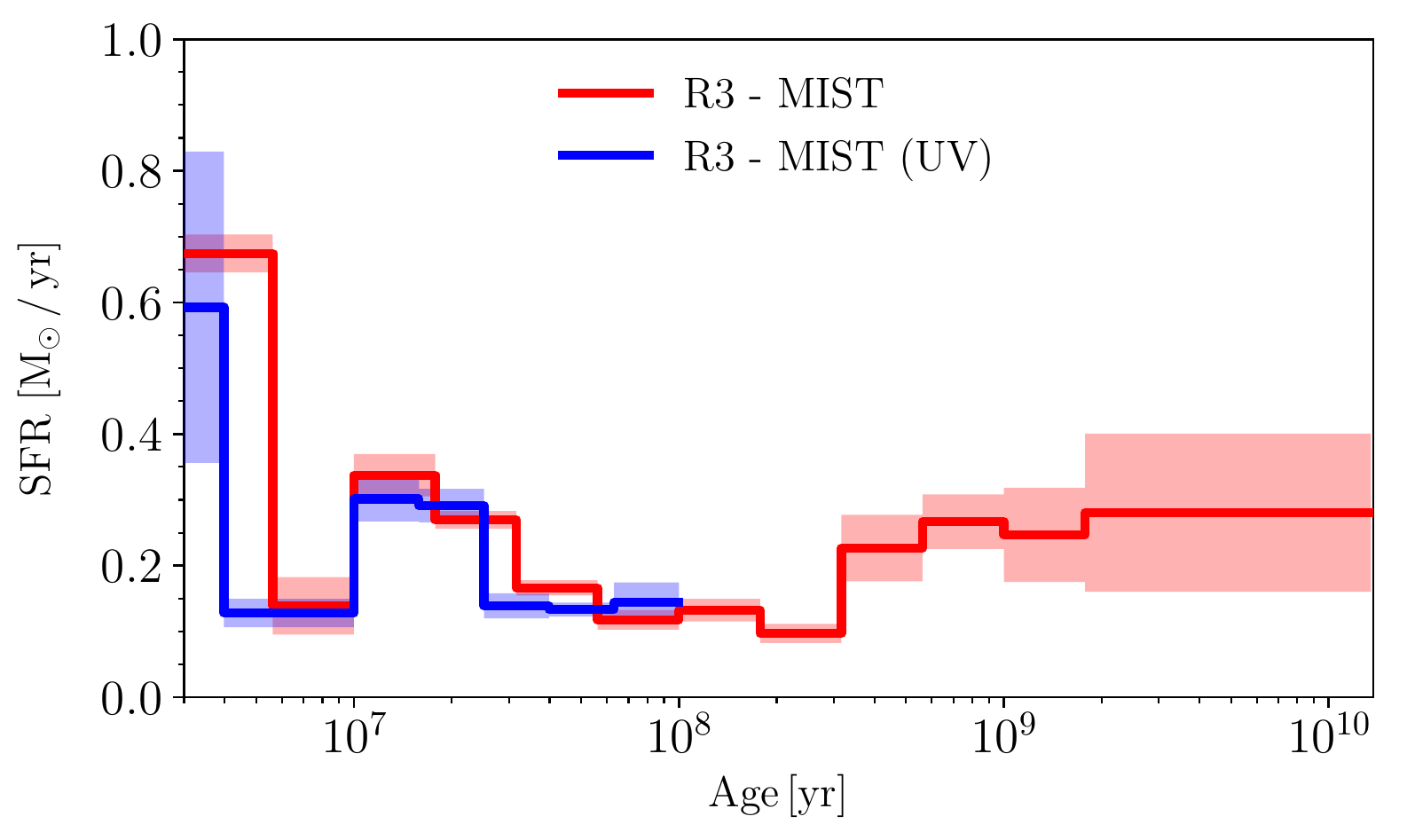}
\caption{Same as Figures \ref{7793_R1} and \ref{7793_R2} but for the third annulus of NGC~7793.}
\label{7793_R3}
\end{figure*}

\section{Star Formation History} \label{sec:7793_sfh}
To recover the SFH from an observational CMD, a very powerful approach is the comparison with synthetic models, as first conceived by \citet{Tosi1991} and then applied by different groups (see, e.g., \citealt{Tolstoy2009} for a review). Our procedure is that described in detail by \citet{Cignoni2015} but implemented with new, updated routines, as illustrated by \citet{Cignoni2018} and \citet{Sacchi2018}. Briefly, the synthetic CMDs are created starting from theoretical evolutionary tracks or isochrones by deriving from them the luminosity and temperature corresponding to the mass and age of the synthetic stars, extracted with a Monte Carlo approach from a random IMF-weighted sample. They are built as linear combinations of ``basis functions'' (BFs), i.e. contiguous star formation episodes whose combination spans the whole Hubble time. Theoretical (luminosity vs temperature) synthetic CMDs need to be converted into the observational (magnitude vs color) plane adopting proper photometric conversions, distance and reddening of the examined galaxy, photometric errors and blends, and incompleteness of the data.

By repeating the process for many metallicities, we end up with a library of synthetic CMDs for different ages and metallicities. The comparison between observed and synthetic CMDs is performed on the so-called Hess diagrams, i.e. the density of points in the CMD. To this purpose, the BFs, as well as the observed CMD, are pixelated into a grid of color and magnitude bins. In each cell, the BFs are linearly combined to match the number of observed stars in that cell:
\begin{equation}
    N(i)=\sum_{Z}\sum_{t} w_{Z,t} \times \mathrm{BF}_{Z,t} \, ,
\end{equation}
where $w_{Z,t}$ are the weights of each BF, and represent the SFR at a given time and metallicity step.

We perform the minimization of the residuals between data and models adopting a Poissonian approach, since we need to consider that some parts of the CMD might have a low number of stars. We implemented the construction of the synthetic CMDs and the comparison between models and data in the hybrid genetic code SFERA \citep{Cignoni2015}.

To build the basis functions for this specific galaxy, we used the information coming from the artificial star tests described in Section \ref{sec:7793_ast}, together with the galaxy's distance modulus, $27.87$ \citep{Radburn-Smith2011,Sabbi2018}, which in our code is allowed to vary from the chosen literature value up to $\pm 0.2$ in steps of 0.05, and metallicity, $Z = 0.0015 - 0.0152$, or $[\mathrm{M/H}]$ from $-2.0$ to $0.0$, adopting the approximation $[\mathrm{M/H}] \simeq \log(Z/Z_{\odot})$. In the most recent $\sim 50$~Myr, the metallicity was allowed to vary in the smaller range $[0.0076-0.0152]$, around the value inferred from spectroscopy of the H~\textsc{ii} regions of the galaxy \citep{Bibby2010}.

To check possible systematic effects due to the adopted stellar evolution models, we derived the SFH using two different sets of models. The models were created from the PARSEC-COLIBRI \citep{Bressan2012,Marigo2017} and MIST \citep{Choi2016} isochrones\footnote{Notice that the PARSEC-COLIBRI models adopt $Z_{\odot} = 0.0152$, while the MIST models adopt $Z_{\odot} = 0.0142$.} using a \citet{Kroupa2001} IMF from $0.1$ to $350$~M$_{\odot}$ and $30\%$ of binary stars.

Extinction is modeled using two parameters: the total extinction, including both foreground and internal, that we varied from 0.05 to 0.20 mag in steps of 0.025, and a differential reddening (modeled as an additional spread) varying from 0 to 0.20 mag. 
The differential reddening of the youngest MS population is modeled separately from that of the rest of the stars \citep{Dolphin2003}. We tried to use different ``shapes'' for the differential reddening modeling (including a Gaussian spread, as in \citealt{Sacchi2018}) with no significant differences or improvements from the adopted one. The same extinction law is adopted for both data sets \citep{Cardelli1989} with $R_{V} = 3.1$. The foreground extinction, consistently measured in the different regions through our CMD-fitting procedure, is $0.07\pm 0.02$, consistent within the errors with the literature values 0.06 from \citet{Schlegel1998} and 0.05 from \citet{Radburn-Smith2012}. 

\subsection{Region 1}
Figure \ref{7793_R1} shows the results obtained for the innermost region of NGC~7793 with the two sets of stellar models. In the top and middle panels we show the observed and recovered CMDs as Hess diagrams, together with the residuals between the two in terms of the likelihood used to compare data and models in SFERA, i.e., $data \times \ln(data/model) - data + model$; the region below the 50\% completeness limit that we used as limiting magnitude to search for the best SFH is marked with the shaded area, that we excluded from the minimization. In the bottom panels, we show the star formation rate as a function of age recovered from the V/I (in red) and U/V (in blue) catalogs using either the COLIBRI (left column) and or the MIST (right column) isochrones. The youngest bin represents the SFR up to the present.

In the V/I case, the synthetic CMDs well reproduce 
\begin{minipage}{\columnwidth}
the observational one, in particular the one based on the COLIBRI isochrones, both in terms of shape and recovered number of stars. However, the bright tip of the red plume (at $\mathrm{m_{F555W}-m_{F814W}} > 2$, $\mathrm{m_{F814W}} < 22.5$), where the red He-burning stars are located, is not well matched by the MIST-based CMD, likely due to a difficulty in modeling this phase of stellar evolution. The whole red plume is indeed not well reproduced by these models, that produce a more vertical, broader, and fainter sequence. In both solutions, the models leave a small gap between the MS and the blue edge of the BL (as already noticed and discussed by \citealt{Cignoni2018}), both in the V/I and U/V cases. This might be caused by a non-optimal modeling of the blue loop color extensions (thus, temperature excursion), that are highly sensitive to metallicity. This kind of study can indeed provide new input information for stellar evolution models to better calibrate and constrain some parts of the CMD, in particular in the shortest wavelength bands that are rarely used in this context.
\end{minipage}

The star formation activity in this central region of the galaxy is mostly older than $1-2$~Gyr, but continues until recent epochs, in agreement with the H$\alpha$ emission that we see in Figure \ref{color_7793}. The two solutions are quite similar, and the differences between the two can be interpreted as an estimate of the systematic uncertainties caused by a different treatment of some -not fully understood- aspects of stellar evolution (e.g. mass loss, convective core overshooting, stellar rotation, atomic diffusion in low-mass stars, bolometric corrections for cool stars, uncertainties around the enrichment law, or the abundance of $\alpha$-elements). Also, the V/I and U/V SFHs in the most recent $\sim 100-200$ Myr well agree, although the U/V SFR is considerably lower than the V/I SFR (derived with the same stellar models) between $\sim 50$ and $100$~Myr. This is the edge of the lookback time reachable with our U/V data, corresponding to CMD regions that start to be much more affected by incompleteness. In fact, the MIST solution does not even reach the same lookback time as the COLIBRI one because of the different treatment of some stellar evolution parameters, so the U/V solutions should be treated carefully in the last time bins.

\begin{figure*}
\centering
\includegraphics[width=0.49\linewidth]{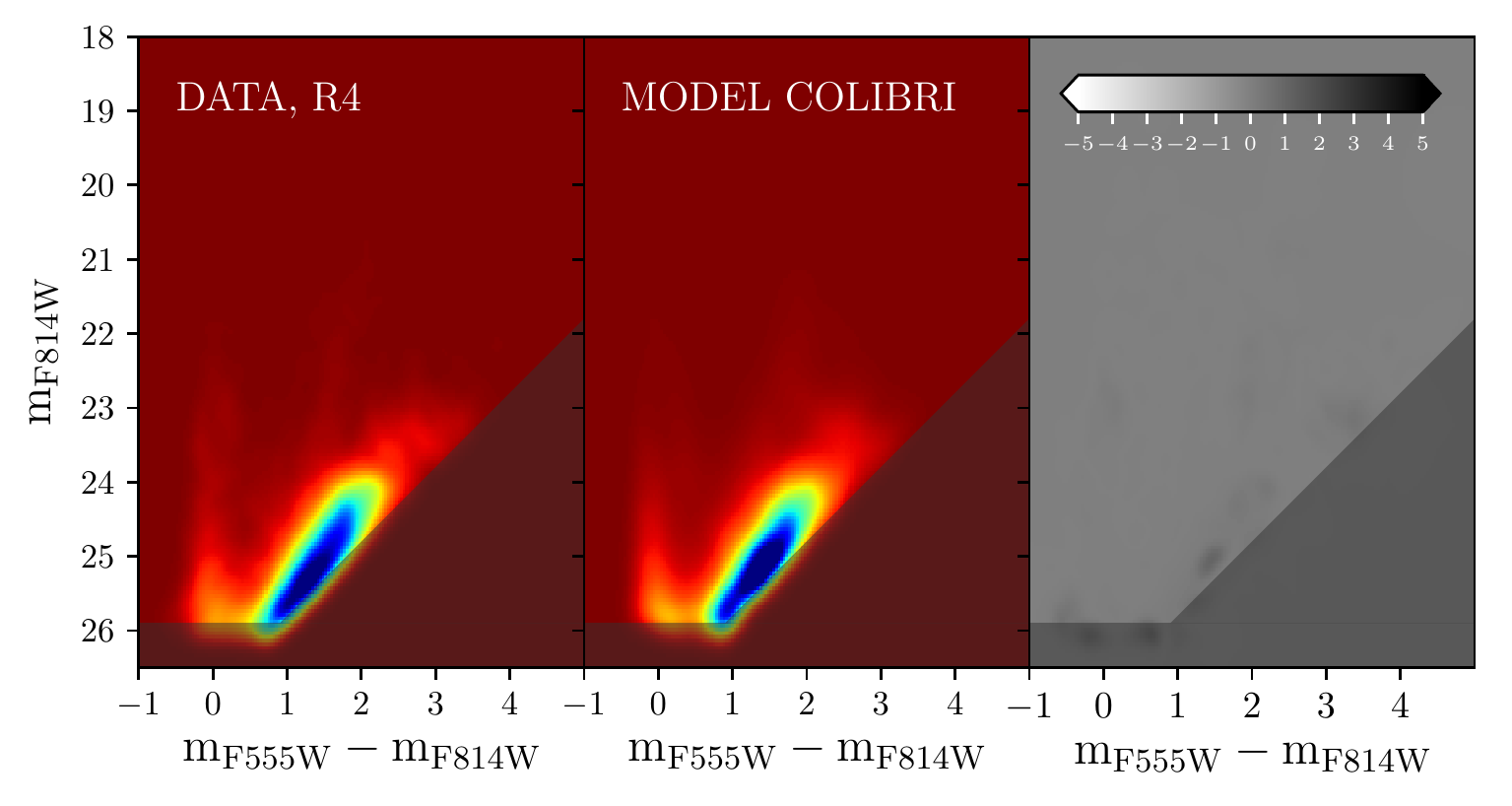}
\includegraphics[width=0.49\linewidth]{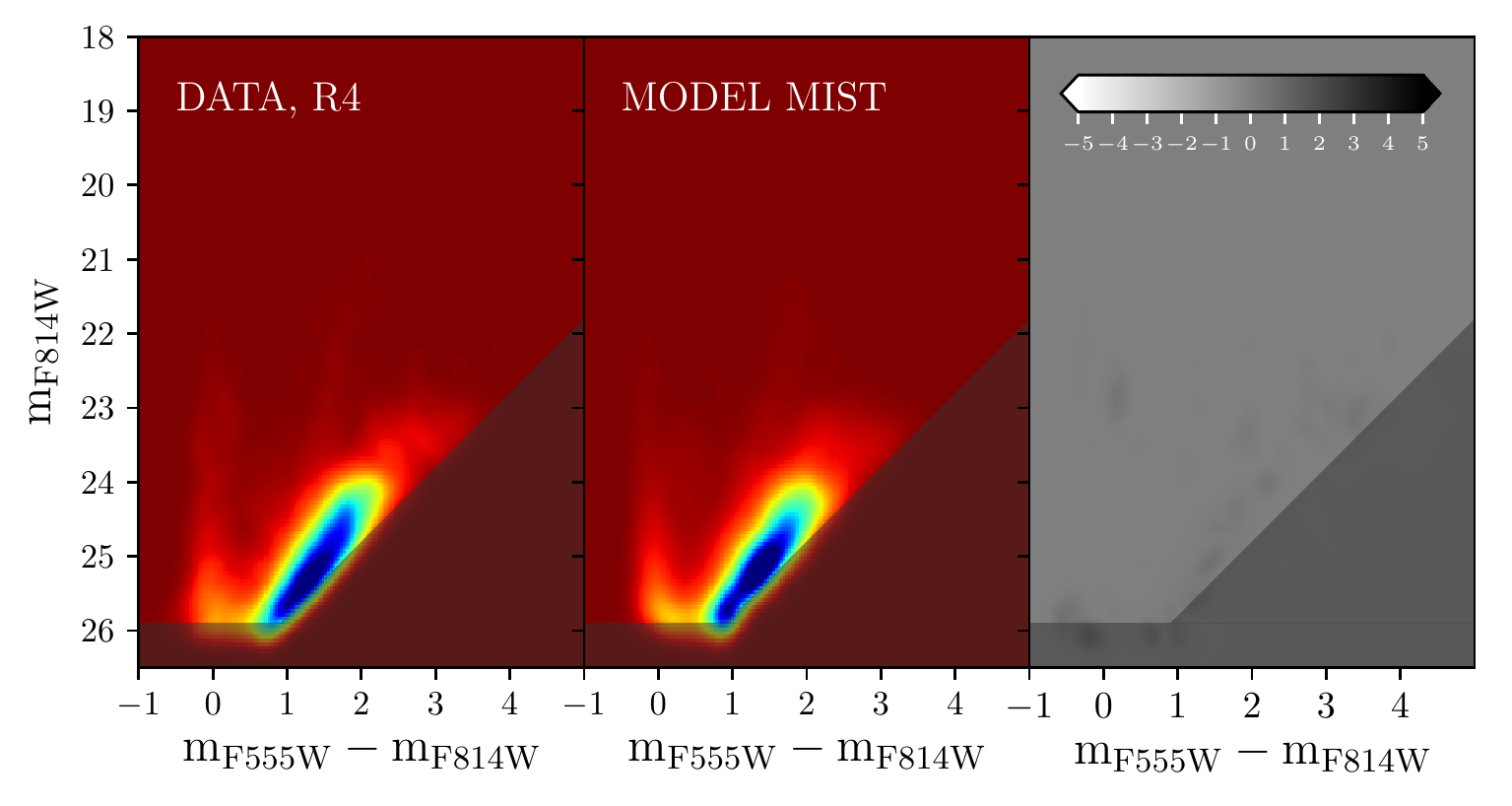}
\centering
\includegraphics[width=0.49\linewidth]{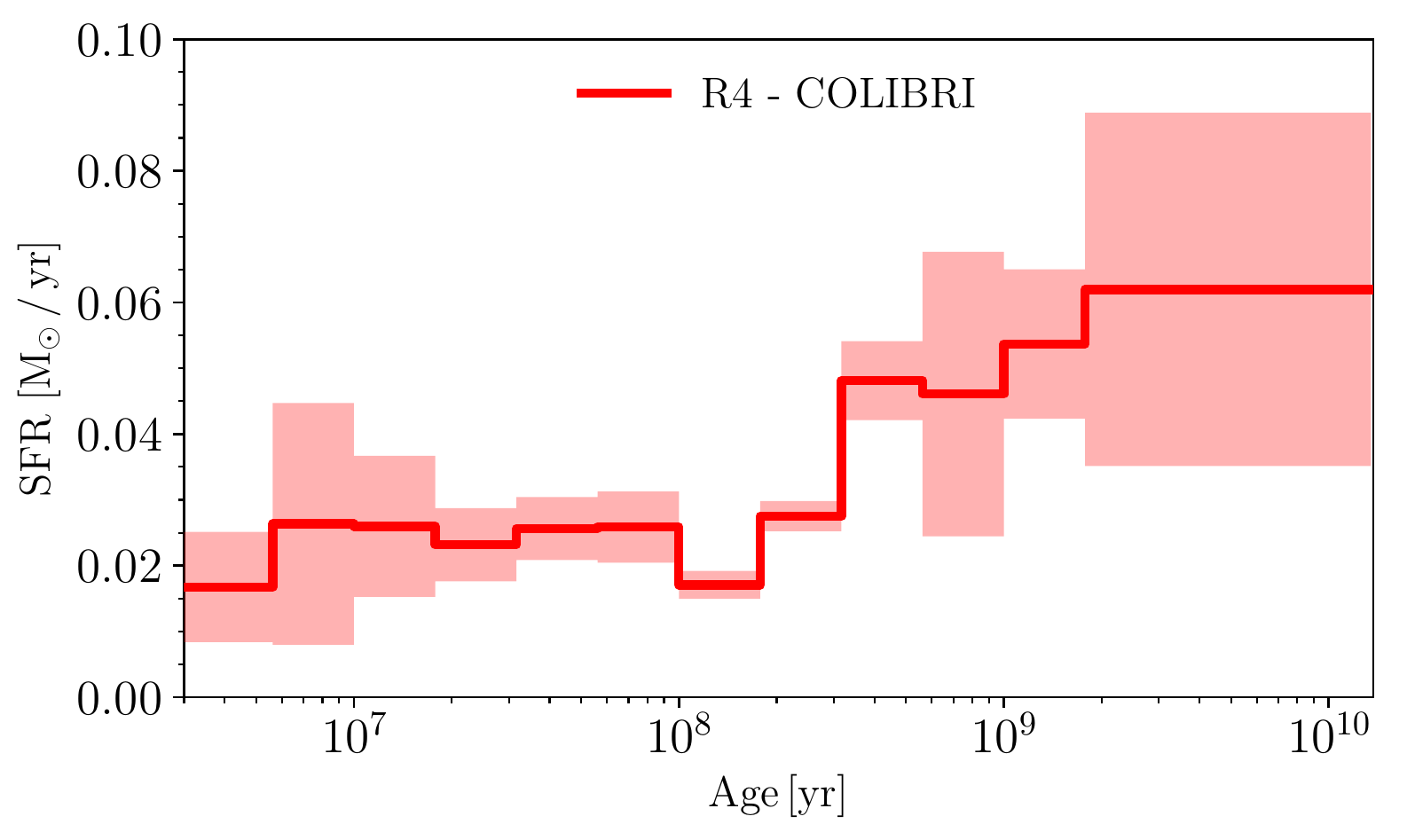}
\includegraphics[width=0.49\linewidth]{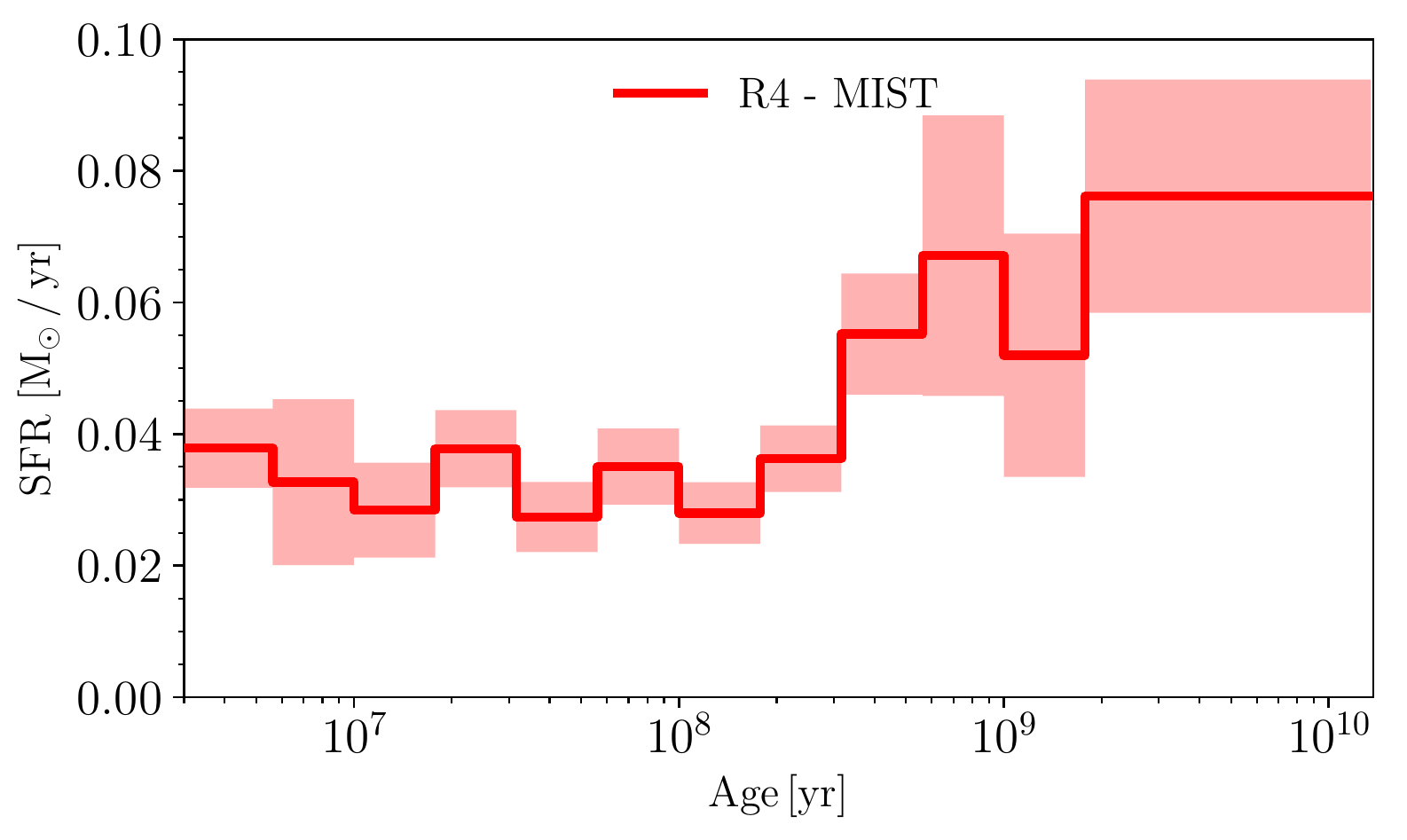}
\caption{Same as Figures \ref{7793_R1} to \ref{7793_R3} but for the outermost region of NGC~7793. Our WFC3/UVIS data do not cover this part of the galaxy, thus we could not recover the U/V SFH here. Notice that the SFR scale is different from that in the previous figures for sake of visibility of the much lower SFRs.}
\label{7793_R4}
\end{figure*}

\begin{table*}[ht]
\caption{Summary of the star formation rates and stellar masses in the different regions of the galaxy from our V/I catalog.\\
$^{(*)}$ The last column is the ratio between the two previous ones.}
\begin{center}
\begin{tabular}{ccccccc}
    \toprule
    \midrule
    \addlinespace[0.3em]
    \multirow{2}{*}{region} & $\mathrm{\langle SFR \rangle}$ & $\mathrm{SFR_{\, peak}}$ & \multirow{2}{*}{$\mathrm{age_{\, peak}}$} & $\mathrm{M_{\ast}(age \leq 50\ Myr)}$ & $\mathrm{M_{\ast}(age > 1\ Gyr)}$ & young/old SFR $^{(*)}$\\
    \addlinespace[0.3em]
    & $\mathrm{[M_{\odot}/yr/kpc^2]}$ & $\mathrm{[M_{\odot}/yr/kpc^2]}$ & & $\mathrm{[10^6\ M_{\odot}]}$ & $\mathrm{[10^9\ M_{\odot}]}$ & [$10^{-3}$]\\
    \midrule
     \multicolumn{7}{c}{\textbf{COLIBRI}}\\
    \midrule
    1 & $0.035 \pm 0.012$ & $0.039 \pm 0.013$ & $2-13.7$ Gyr & $1.23 \pm 0.08$ & $1.10 \pm 0.23$ & 1.1 \\
    2 & $0.011 \pm 0.003$ & $0.022 \pm 0.001$ & $0-6$ Myr & $2.50 \pm 0.11$ & $0.85 \pm 0.16$ & 2.9 \\
    3 & $0.008 \pm 0.003$ & $0.014 \pm 0.001$ & $0-6$ Myr & $2.85 \pm 0.14$ & $0.83 \pm 0.17$ & 3.4 \\
    4 & $0.005 \pm 0.002$ & $0.005 \pm 0.002$ & $2-13.7$ Gyr & $0.27 \pm 0.04$& $0.15 \pm 0.04$ & 1.8 \\
    \midrule
     \multicolumn{7}{c}{\textbf{MIST}}\\
    \midrule
    1 & $0.029 \pm 0.012$ & $0.033 \pm 0.027$ & $2-13.7$ Gyr & $1.40 \pm 0.07$ & $0.91 \pm 0.37$ & 1.5 \\
    2 & $0.009 \pm 0.005$ & $0.021 \pm 0.002$ & $0-6$ Myr & $2.71 \pm 0.08$ & $0.72 \pm 0.38$ & 3.7 \\
    3 & $0.007 \pm 0.003$ & $0.013 \pm 0.001$ & $0-6$ Myr & $2.97 \pm 0.10$ & $0.70 \pm 0.28$ & 4.3 \\
    4 & $0.006 \pm 0.001$ & $0.006 \pm 0.003$ & $2-13.7$ Gyr & $0.35 \pm 0.04$ & $0.19 \pm 0.04$ & 1.9 \\
    \midrule
    \bottomrule
\end{tabular}
\end{center}
\label{table-7793}
\end{table*}

\subsection{Region 2}
Figure \ref{7793_R2} shows the results of the SFH recovery process for the middle ring-shaped region of the galaxy (green, in Figures \ref{7793_fields} and \ref{7793_fields_uv}). The observational CMD exhibits a much more populated MS with respect to that of the inner region, as expected from the presence of many star forming regions and H$\alpha$ emission in the spiral arms. Indeed, the recent SFR is higher ($\sim3$ times more, in the most recent bin) than that of Region 1, and the ratio between old and young SF is much lower. There is a general good agreement between the solutions from the two sets of models.
The CMD from the COLIBRI solution well matches the data, with the only caveat that the model produces a red plume broader than the data, at $\mathrm{m_{F814W}} < 22.5$ and $\mathrm{m_{F555W}-m_{F814W}} \gtrsim 1.5$. The MIST solution has similar performances, with a worse match of the red plume, which is fainter and broader than the observed one as we already found in Region 1, suggesting a systematic effect in the models. As in the inner Region, again the MIST CMDs show the gap between the blue edge of the blue loops and the MS, due to a too short color extension of the loops.
\\

The U/V SFH is systematically lower than the V/I one, even though they agree within the 1$\sigma$ errors except for the time bins between $\sim 50$ and $100$~Myr in the COLIBRI solution and between $\sim 20$ and $50$~Myr in the MIST solution. We ascribe these large differences mainly to systematics from the stellar evolution libraries: the morphology of the U/V stellar models still have some issues, e.g. the MS and He-burning stars of intermediate mass tend to have a gap in the models that is not seen in the data. Moreover, extinction is a bigger challenge in U/V CMDs; for instance, we used the same extinction law for both data sets, while there is no general consensus on whether this is correct at shorter wavelengths, in particular in low metallicity galaxies. Here and in Region 3 there is the additional caveat of the different coverage of the ACS (used for the V/I data) with respect to the WFC3 (used for the U/V data) leading to two effects: the sampled area is a bit different (smaller for the U/V, see Figure \ref{color_7793}), and some star forming regions might have been included in the ACS V/I field catalog but not in the WFC3 U/V one, possibly giving rise to these differences.\\

\subsection{Region 3}
Figure \ref{7793_R3} shows the CMDs and SFHs for Region 3 of NGC~7793. In the COLIBRI case the observational CMD features are well reproduced by the synthetic ones, with the only exception of the red edge of the TP-AGB stars, that produce a feature with a color hard to model. The SFH is generally very similar to that of Region 2, and also the U/V SFHs show a very similar trend, with a general good agreement with the V/I ones, in both solutions.

As in the previous region, the MIST solution shows more noisy residuals, it produces a more evident gap between MS and BL, and it underestimates the brightest red He-burning stars. In both cases the ratio between recent and ancient SF is slightly higher than in the two more internal regions, confirming the trend we already noticed that suggests an inside-out growth of the galaxy.

\subsection{Region 4}
This region includes a very small area of the galaxy, that was covered by the ACS only, thus we could not recover the U/V SFH here. As shown in Figure \ref{7793_R4}, the V/I CMD presents a faint MS, some He-burning and TP-AGB stars, and the RGB. The SFH smoothly decreases from the oldest epochs to the present, in both cases. Even though this might look like an inversion of the trend we find in the three regions previously analyzed, we need to use caution when comparing this region to the others, given the small portion of the galaxy covered by it (see Figure \ref{7793_fields}). The agreement between observational and synthetic CMDs is very good, with no relevant discrepancies in both solutions.\\

A summary of the SFRs and stellar masses formed at various epochs in the different regions of the galaxy is given in Table \ref{table-7793}.


\begin{figure*}[!ht]
\centering
\includegraphics[width=0.50\linewidth]{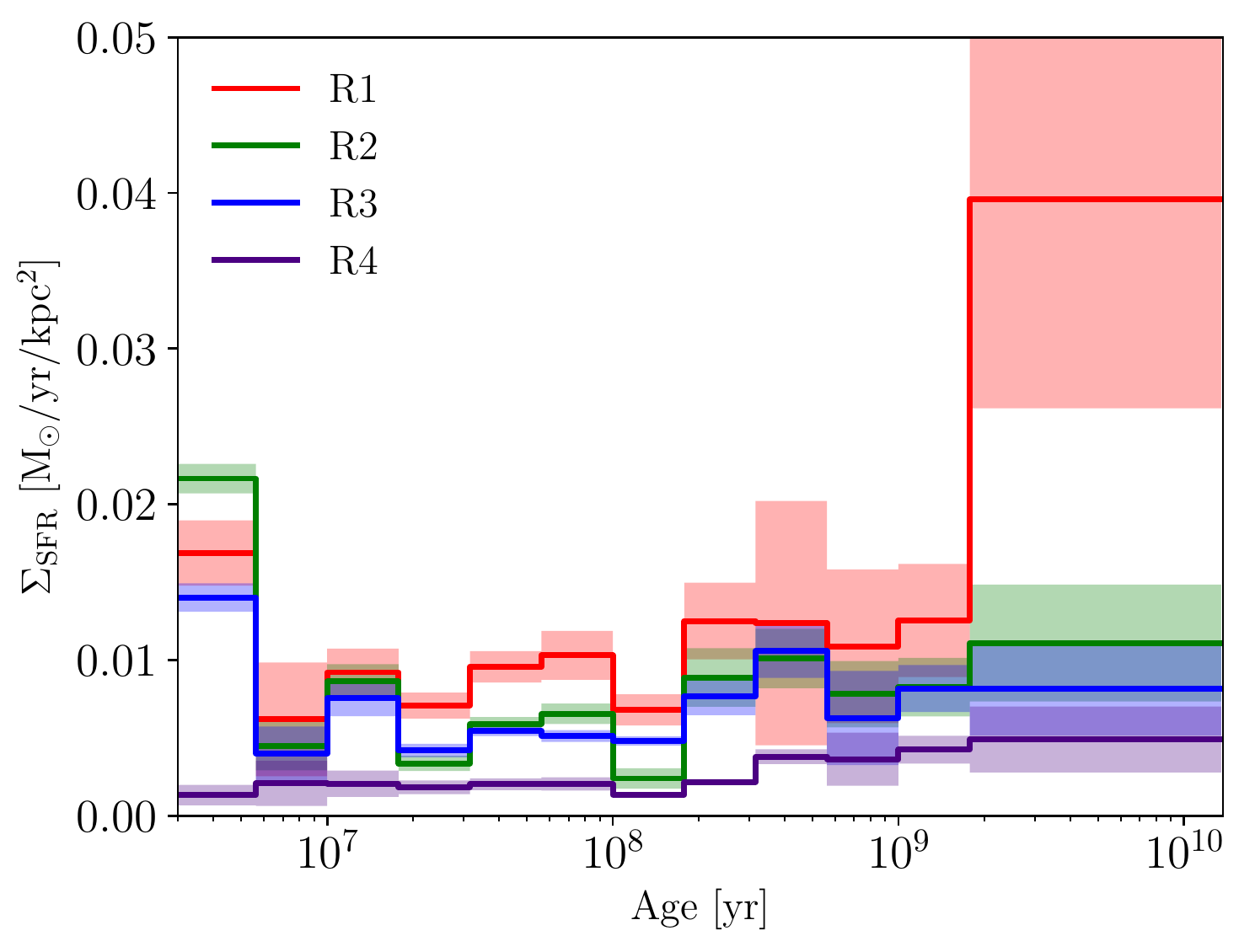}
\includegraphics[width=0.49\linewidth]{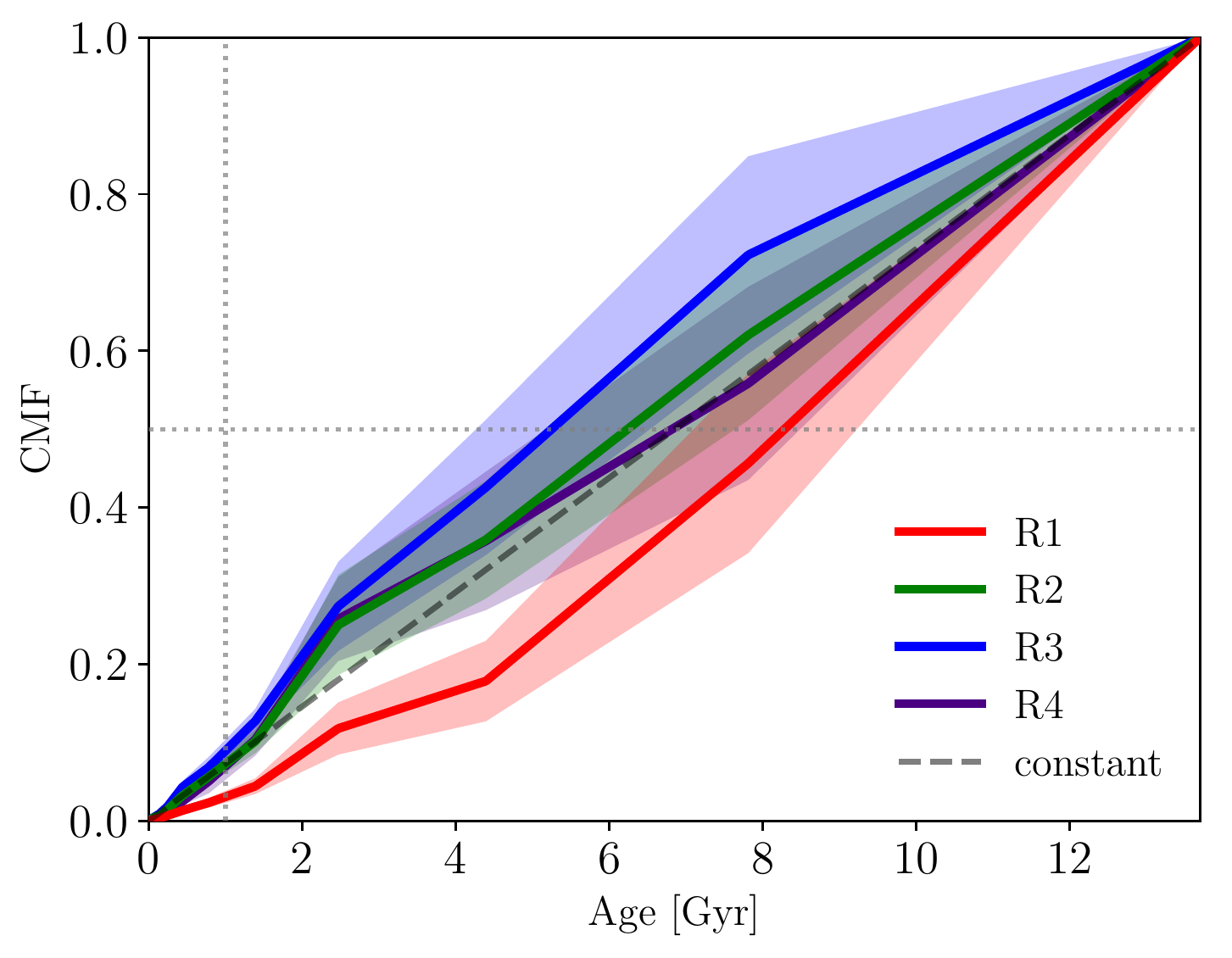}
\caption{Optical SFR surface densities (SFR/area, left panel) and cumulative stellar mass fraction (right panel) as a function of time in different regions of NGC~7793, from the COLIBRI solutions. The right hand plot also shows a comparison with a constant SFH (black solid line), and a reference for an age of 1 Gyr (vertical dotted line) and for 50\% of the total assembled mass (horizontal dotted line).}
\label{7793_all-density}
\end{figure*}

\section{Discussion}
Spiral galaxies have a major role in the universe today, with disk galaxies dominating star formation in the low-redshift universe \citep{Brinchmann2004}.  Spiral galaxies have also been shown to operate at the peak baryon efficiency, more readily converting their gas into stars than galaxies at higher and lower masses \citep{Guo2010}. In addition, they provide an important laboratory for studying star formation due to the synchronization of SF by density waves, which is essential to determine the timescales in the cloud formation $-$ star formation $-$ cloud disruption cycle. Understanding the processes that convert raw molecular material into stars and how star formation feeds back into galaxies and their environments in spiral galaxies is of great interest.

The spiral structure and strength of the density waves may affect the molecular cloud properties in a variety of ways. For example, simulations find that the longest lived clouds tend to be the most massive and bound, with these clouds surviving into the inter-arm regions and containing relatively older massive clusters \citep{Dobbs2014}. Clouds in the spiral arms, where they are likely more affected by feedback, tend to be destroyed (and have shorter lifetimes), or merge to form more massive clouds.  If the galaxy is a grand design spiral, then the clouds may show a sequence across the spiral arms, with clouds forming from H$_2$, then stars and clusters forming in complexes or large structures \citep{Gouliermis2017,Grasha2017b}. In flocculent galaxies, a more random distribution of clouds and evolution is expected \citep{Dobbs2010}, but age spreads in simulations are found to be even shorter than in grand design galaxies \citep{Dobbs2014}.

In this paper, we presented the spatially resolved SFH of the central part of NGC~7793, a face-on flocculent spiral galaxy which we studied in the F336W, F555W and F814W filters thanks to archival and new \textit{HST} data within the LEGUS Treasury program. The galaxy has also been re-observed with ALMA (PI: Kelsey Johnson) in order to study the impact of spiral structure and feedback from stellar populations on molecular clouds, and the spatial correlation between the star clusters and molecular gas. Figure 2 of \citet{Grasha2018} shows the UVIS F438W image of our two pointings of NGC~7793, overlaid with the giant molecular clouds, star clusters, and the outline of the ALMA coverage. It can be easily seen how younger star clusters are substantially closer to molecular clouds than older star clusters, and, as expected, preferentially located on the spiral arms.

In order to ease the comparison among the different sub-regions of NGC~7793, Figure \ref{7793_all-density} shows the SFHs of the analyzed regions normalized to the corresponding area (left panel) and the cumulative stellar mass fraction formed at different epochs (right panel). Since we cover only the central portion of the galaxy, these SFHs are very similar once normalized to their area, but the main difference among them is that the ratio of present-to-past averaged star formation rate increases from Region 1 to Region 3 (see also Table \ref{table-7793}), suggesting an ``inside-out'' growth scenario for the disk of NGC~7793, a mechanism originally introduced in the theory of galaxy evolution on the basis of chemical evolution arguments \citep[see, e.g.,][]{Larson1976,Matteucci1989}, and used as a basis for semi-analytic modeling of disk galaxies in the context of cold dark matter cosmologies \citep{Kauffmann1996,vandenBosch1998}. Under the assumption of detailed angular momentum conservation, the inside-out picture reflects the distribution of specific angular momentum of the protogalaxy: gas accreted at late times has a higher specific angular momentum and settles in the outer regions of the galaxies, therefore the outskirts of spiral galaxies are expected to form later. However, hydrodynamical simulations of the formation of disk galaxies indicate that angular momentum is not conserved and disks do not always form from the inside out \citep{Sommer-Larsen2003}. Whatever the arguments about angular momentum, a higher accretion of metal-poor gas and/or a longer e-folding time of the accretion rate as a function of the galactocentric distance is  needed to allow for the metallicity gradients observed in spiral galaxies (e.g. \citealt{Tosi1988}, \citealt{Matteucci1989}, \citealt{Magrini2007}, \citealt{Minchev2014}, \citealt{Kubryk2015}).

Another reason for this inside-out growth could also be just a depletion of gas in the inner parts, by consumption and turning into stars. The density is higher in the inner parts and the consumption time (for total gas, not just the observationally selected molecules) smaller, thus the inner parts run out first. There could also be continued accretion in the outer parts, and so both processes probably work together. The shorter consumption time for inner disk gas follows from the steep slope of the Kennicutt-Schmidt (KS) relation for the total gas \citep{Elmegreen2015}.

This inside-out behavior can be better visualized by using the cumulative stellar mass fraction (right panel of Figure \ref{7793_all-density}), that shows how SF in Region 1 is below a constant SFR and proceeded much faster at older epochs with respect to most recent ones. The opposite happened in Regions 2 and 3, whose SFRs are above a constant one, with slopes indicating a faster SF in the last few Gyr. In particular, Region 1 had already formed 50\% of its total stellar mass around $\sim 8.3$~Gyr ago, while this occurred $\sim 6.2$~Gyr ago for Region 2, and $\sim 5.2$~Gyr ago for Region 3 (see the horizontal dotted line for a reference). Interestingly, Region 4 seems to deviate from this behavior, with much less populated MS and He-burning phases with respect to Regions 1-3 and a prevalence of older stars (see Figure \ref{7793_fields}), despite being the more external part of the galaxy we analyzed. This might indicate that the inside-out star formation trend does not progress radially beyond Region 3. However, it is quite evident from Figure 1 that we are not sampling the star forming regions present on the east side of Regions 3 and 4, due to the smaller coverage of the WFC3 and the position of the fields. We thus believe that the young SF is underestimated in the two more external regions, even though this might not be enough to extend the inside-out behaviour to Region 4.

\begin{figure}
\centering
\includegraphics[width=\linewidth]{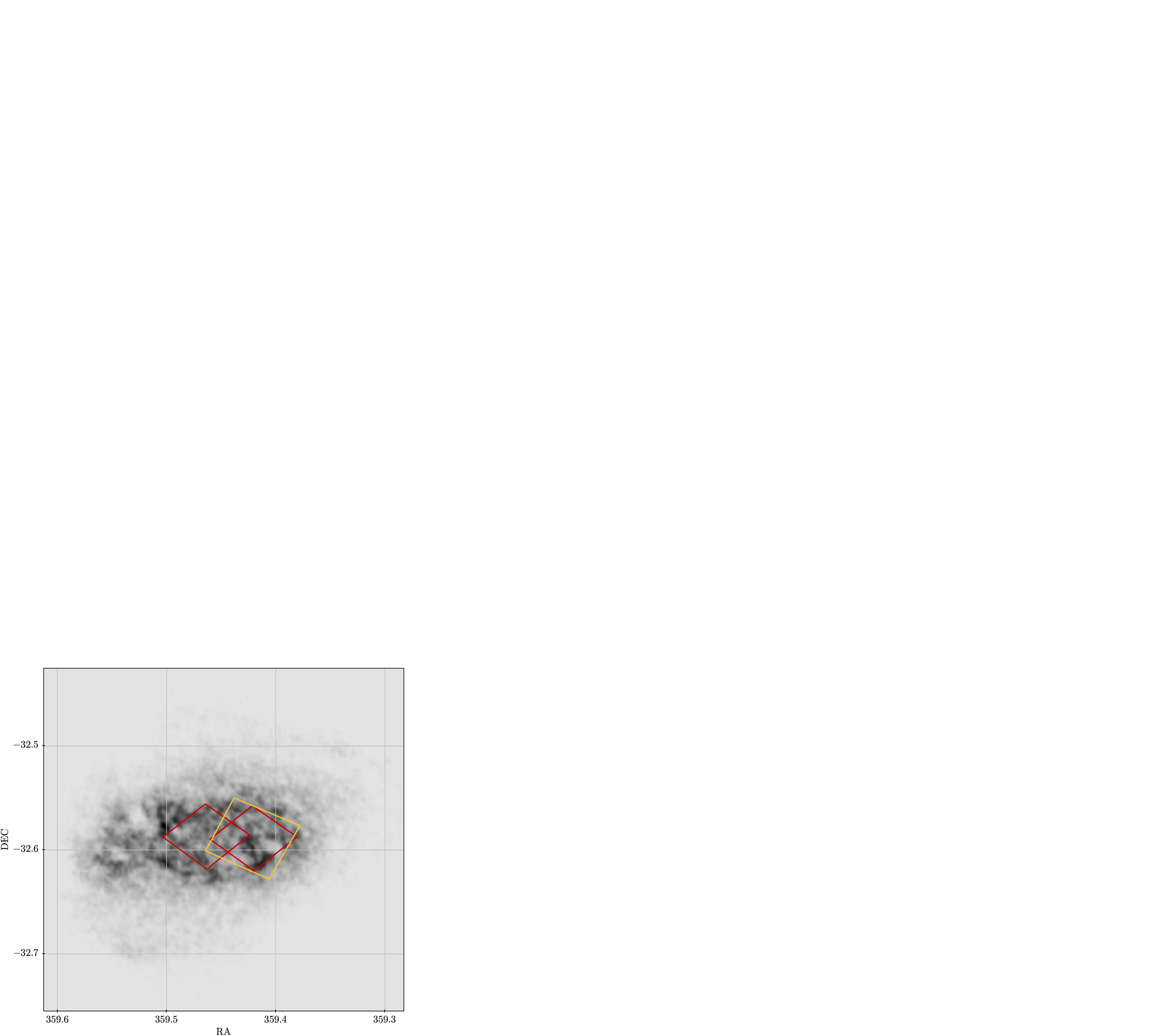}
\caption{Integrated H~\textsc{i} map of NGC~7793 ($\sim 20\times 20$ arcmin) derived from the THINGS data cube \citep{Walter2008}. The gray scale range is from 0 to 507 Jy/beam km/s. Overplotted, the LEGUS UVIS (red) and ACS (orange) footprints that show the coverage of our data compared to the H~\textsc{i} disk of the galaxy.}
\label{7793_hi}
\end{figure}

From an observational point of view, there are several evidences of the inside-out mechanism. \citet{Pohlen2006} studied a sample of 90 local disk galaxies using imaging data from the SDSS survey and found that 60\% of them have inner exponential profiles followed by a steeper outer exponential profile, while 30\% have a shallower (upward bending) outer profile. \citet{Azzollini2008} found that for a given stellar mass, the radial position of this break has increased with cosmic time by a factor of $1.3 \pm 0.1$ between $z = 1$ and $0$, suggesting a moderate inside-out growth of disk galaxies over the last $\sim 8$~Gyr. Using the color profiles of 86 face-on spiral galaxies, \citet{deJong1996} found that the outer regions of disks are on average younger and have lower metallicity. \citet{Munoz-Mateos2007} studied specific SFR (i.e. SFR per unit of galaxy stellar mass) profiles of a sample of 160 nearby spiral galaxies from the GALEX atlas of nearby galaxies \citep{GildePaz2007} and found a large dispersion in the slope of the specific SFR profiles, with a slightly positive mean value, which they interpreted as implying moderate net inside-out disk formation. Moreover, studying the relationship between age, metallicity, and $\alpha$-enhancement of FGK stars in the disk of our Galaxy, \citet{Bergemann2014} found older, more $\alpha$-rich, and more metal-poor stars in the inner disk, hence supporting the inside-out mechanism. Finally, \citet{Pezzulli2015} measured the instantaneous mass and radial growth of the stellar disks of a sample of 35 nearby spiral galaxies that includes NGC~7793, finding clear signatures of its inside-out growth. Within the Local Group, the spatially resolved star formation history analyses of M31 from the PHAT survey (e.g., \citealt{Williams2017}) and M33 \citep{Barker2007,Barker2011} have shown similar behaviors.

Using our results on the SFH, and with literature values for the gas component, we estimated the $\Sigma_{\mathrm{SFR}}$ and $\Sigma_{\mathrm{gas}}$ for NGC~7793. This is particularly useful to place the galaxy on the KS law \citep{Schmidt1959,Kennicutt1989}, an observational correlation between the two quantities. In principle, one should measure the total gas quantity to address the correct relation; in practice, it is not always trivial to obtain information about the molecular component, thus, in some cases we need to rely on the H~\textsc{i} alone. For NGC~7793, we used the data from THINGS \citep{Walter2008} to infer the H~\textsc{i} mass within the LEGUS footprints (see Figure \ref{7793_hi}), obtaining $\mathrm{M_{H~\textsc{i}} = 2.54 \times 10^8}$~M$_{\odot}$, corresponding to $\log(\Sigma_{\mathrm{gas}}/\mathrm{M_{\odot} pc^{-2}}) = 1.29$, after adopting the conversion $\Sigma_{\mathrm{gas}} = 1.36\ (\Sigma_{\mathrm{H~\textsc{i}}}+\Sigma_{\mathrm{H_2}})$ to take into account the presence of helium. From our SFH in the last 10~Myr, we obtain $\log(\Sigma_{\mathrm{SFR}}/\mathrm{M_{\odot} yr^{-1} kpc^{-2}}) = -1.99$, placing NGC~7793 on the KS relation. This analysis could be further improved by using volume densities instead of surface-based quantities, as proposed and investigated by \citet{Bacchini2019}.

One interesting question is whether spiral arms trigger star formation, or whether they simply ``rearrange'' young stars, or molecular clouds in the galaxy. \citet{Roberts1969} considered the response of gas to stellar spiral arms and showed that the gas experiences a strong shock. The sharp rise in density naturally means that the densities required to produce molecular gas, and gravitational collapse to form stars are reached, so the idea that spiral arms trigger star formation was proposed. However various observational results queried spiral arm triggering of star formation. \citet{Elmegreen1986} compared the star formation rates in flocculent and grand design galaxies, and found that there was little difference despite the different spiral arms. More recent observations still debate this: \citet{Eden2013} and \citet{Foyle2010} find little difference in the star formation efficiencies in spiral arms and inter-arm regions,while \citet{Seigar2002} do find a dependence of star formation rate on the strength of spiral arms. The spatial distribution of old ($> 1$~Gyr) stars in NGC~7793 (presented in Figure \ref{7793_pop}), suggested there are no perceptible spiral density waves in this galaxy, and that the spiral structures we see in visible and UV light are simply star forming regions, not spiral waves.

The comparison of our results for NGC~7793 with other analyses of different spiral galaxies will contribute to this debate, hopefully helping to understand the role and impact of spiral arms on the triggering and efficiency of star formation.


\section{Summary and Conclusions}
Here we summarize the main results of this paper, that analyzed for the first time the resolved stellar populations and radial SFH of this spiral galaxy outside the Local Group.

\begin{itemize}
    \item[-] From both archival and new LEGUS data, we determined an average SFR over the whole Hubble time of $0.23\pm0.02$~$\mathrm{M_{\odot}/yr}$, corresponding to a total stellar mass of $(3.09\pm0.33)\times 10^9$~$\mathrm{M_{\odot}}$.
    \item[-] The new F336W photometry allowed us to better separate MS and post-MS stars in the CMD, so to analyze the youngest stellar populations with a higher color and time resolution.
    \item[-] From the SFH recovered in different radial regions of the galaxy we found a growing trend of the present-to-past SFR ratio, increasing from internal to more external regions, that suggests an inside-out growth of the galaxy (even though it is not clear from our data whether this trend extends to the outer parts of the disk).
    \item[-] Using literature values for the gas component, we estimated $\log(\Sigma_{\mathrm{gas}}/\mathrm{M_{\odot} pc^{-2}}) = 1.29$, and from our SFH in the last 10~Myr we obtained $\log(\Sigma_{\mathrm{SFR}}/\mathrm{M_{\odot} yr^{-1} kpc^{-2}}) = -1.99$, placing NGC~7793 well within the scatter of the KS relation.
    \item[-] The analysis of the spatial distribution of different stellar populations within the galaxy indicated the possible lack of spiral density waves in NGC~7793.
\end{itemize}

\acknowledgments{
These data are associated with the \textit{HST} GO Program 13364 (PI D. Calzetti). Support for this program was provided by NASA through grants from the Space Telescope Science Institute. M.C. and M.T. acknowledge funding from the INAF PRIN-SKA 2017 program 1.05.01.88.04. A.A. acknowledges the support of the Swedish Research Council, Vetenskapsr{\aa}det and the Swedish National Space Agency (SNSA). We really thank the anonymous referee for the uselful comments and suggestions, that produced significant improvements in the present paper.
}

\bibliography{bib}

\end{document}